\documentstyle[epsf]{mn}   

\def\GS{\lower.5ex\hbox{$\buildrel>\over\sim$}}
\def\LS{\lower.5ex\hbox{$\buildrel<\over\sim$}}

\title{Ionisation Structure in Accretion Shocks 
   with a Composite Cooling Function}
\author[K. Wu, M. Cropper \& G. Ramsay]
   {Kinwah Wu$^{1,2}$, Mark Cropper$^2$ \& Gavin Ramsay$^2$ \\ 
$^1$ Research Centre for Theoretical Astrophysics, 
     School of Physics A28, University of Sydney, NSW 2006, Australia \\ 
$^2$ Mullard Space Science Laboratory, University College London, 
     Holmbury St.~Mary, Dorking, Surrey RH5~6NT } 
     
\date{Received: } 

\begin{document}

\maketitle 

\begin{abstract}  

We have investigated the ionisation structure 
  of the post-shock regions of magnetic cataclysmic variables 
  using an analytic density and temperature structure model 
  in which effects due to bremsstrahlung and cyclotron cooling 
  are considered. 
We find that in the majority of the shock-heated region 
  where H- and He-like lines of the heavy elements are emitted, 
  the collisional-ionisation and corona-condition approximations  
  are justified.  
We have calculated the line emissivity and ionisation profiles  
  for Iron as a function of height within the post-shock flow. 
For low-mass white dwarfs, line emission takes place near the shock.  
For high-mass white dwarfs, most of the line emission takes place 
  in regions well below the shock  
  and hence it is less sensitive to the shock temperature. 
Thus, the line ratios are useful to determine the white-dwarf masses 
  for the low-mass white dwarfs, 
  but the method is less reliable when the white dwarfs are massive.   
Line spectra can, however, be used 
  to map the hydrodynamic structure of the post-shock accretion flow.      

\end{abstract}

\begin{keywords} 
accretion ~---~ shock waves ~---~ X-rays: stars ~---~ 
stars: white dwarfs ~--~ methods: data analysis  
\end{keywords} 

\section{Introduction} 

When a supersonic accretion flow abruptly becomes subsonic 
  near the surface of a compact star, 
  a stand-off shock is formed. 
For the white dwarfs in magnetic cataclysmic variables (mCVs) 
  (see Warner 1995), 
the accretion-shock temperature 
  $T_{\rm s} = 3~G \mu m_{_{\rm H}} 
     M_{\rm w}/ 8 k R_{\rm w}$ (if $x_{\rm s} << R_{\rm w}$), 
  where $G$ is the gravitational constant, 
  $k$ the Boltzmann constant, 
  $m_{_{\rm H}}$ the hydrogen mass, 
  $\mu$ the mean molecular weight of the gas, 
  $x_{\rm s}$ the shock height, $M_{\rm w}$ the mass of the white dwarf and
  $R_{\rm w}$ the radius of the white dwarf. 
(Typically, $T_{\rm s} \sim 10^8$~K.) 
 As $T_{\rm s}$ depends on 
  $M_{\rm w}$ and $R_{\rm w}$,    
  and is insensitive to other orbital parameters, 
  the white-dwarf mass can be determined  
  if $T_{\rm s}$ is measured (Rothschild et al.\ 1981; 
  Kylafis \& Lamb 1982, Ishida 1991; Fujimoto \& Ishida 1997; 
  Cropper, Ramsay \& Wu 1998).  

The distance from the shock front to the white-dwarf surface 
  (i.e. the shock height) and 
  the structure of the shock-heated emission region 
  depends not only by $T_{\rm s}$ but also the cooling processes. 
For mCVs, bremsstrahlung and cyclotron cooling are the most important. 
When the accretion rate of the system is high 
  ($\GS\ 10^{16}~{\rm g}~{\rm s}^{-1}$) 
  and the white-dwarf magnetic field is weak ($\sim$ 1~MG), 
  bremsstrahlung cooling usually dominates. 
If the accretion rate is low  
  ($\LS\ 10^{16}~{\rm g}~{\rm s}^{-1}$) 
  and the magnetic field is strong  ($\GS$ 10~MG), 
  then cyclotron cooling is the dominant  process,  
  at least near the shock 
  (Lamb \& Masters 1979; King \& Lasota 1979). 
Line cooling is generally unimportant, 
  except at the bottom of the post-shock region, 
  where the electron temperature $T_{\rm e}$ falls well below $10^7$~K 
  (e.g.\ Kato 1976; Mewe, Gronenschild \& van den Oord 1985).   

It is straightforward to calculate the post-shock accretion flow 
  with only optically thin bremsstrahlung cooling 
  (see e.g. Aizu 1973; Chevalier \& Imamura 1982). 
The inclusion of cyclotron cooling complicates the calculations 
  because of large opacity effects. 
An exact treatment of cyclotron cooling requires the solving 
  of the radiative-transfer and hydrodynamic equations simultaneously, 
  and it is unlikely that simple analytic solutions can be obtained. 
However, for parameters typical of mCVs,  
  the total cooling process can be roughly approximated 
  by a composite cooling function 
  with power laws of density and temperature  
  (Langer, Chanmugam \& Shaviv 1982; 
  Wu 1994; Wu, Chanmugam \& Shaviv 1994).
With this composite cooling function,  
  the hydrodynamic equations governing the post-shock flow  
  can be solved, yielding a closed-form solution.  
With the density and temperature structures 
  of the emission region specified, 
  the X-ray emission from the system can be calculated. 
(See Wu 2000, for a review of accretion shocks in mCVs.) 

X-rays emitted from regions 
  in which both cyclotron and  bremsstrahlung cooling 
  are important are generally softer than 
  those from regions with bremsstrahlung cooling only. 
Moreover, the spectra of the cyclotron-bremsstrahlung cooling shocks
  are richer in emission lines 
  (Cropper, Ramsay \& Wu 1998; Tennant et al.\ 1998).  
Emission lines can also be used 
  to diagnose the density and temperature structures 
  of the shock-heated region 
  (see Fujimoto \& Ishida 1995; Ezuka \& Ishida 1999). 
The grating instruments 
  on board the current X-ray satellites {\it XMM-Newton} and {\it Chandra}  
  can easily resolve many of the emission lines in the $\sim 1$~keV range.
With the spectral resolution of these satellites  
  we will be able to measure the accretion flow 
  very close to the white-dwarf surface.   

From the X-ray continuum, 
  the shock temperature can be determined, 
  and hence the white-dwarf mass 
  (e.g.\ Ishida 1991; Cropper, Ramsay \& Wu 1998; Ramsay 2000; 
  Beardmore, Osborne \& Hellier 2000). In analysing X-ray data, 
  model continua spectra are fitted to the data 
  in order to determine the system parameters.  
As noted above, 
  numerical calculations with exact treatment of cyclotron cooling 
  will in principle produce more accurate model spectra, 
  but its heavy demand on computation time makes it impractical 
  in most situations.  
Semi-analytic methods that can yield reasonably accurate results 
  are therefore more useful in the data analysis 
  to extract system  parameters, such as the shock temperature. 
In this study 
  we investigate the ionisation structures and the line emission 
  of the post-shock emission regions of mCVs 
  in terms of the analytic treatment 
  of Wu (1994) and Wu, Chanmugam \& Shaviv (1994).   
We ignore the effects 
  of  gravity over the height of the shock  
  (e.g.\  Cropper et al.\ 1999) and  
  unequal ion and electron temperatures 
  (e.g.\ Imamura et al.\ 1987; Saxton 1999; Saxton \& Wu 1999).    
      
\section{Structured Shock-heated Regions}   

The ionisation of the elements in the post-shock flow in mCVs
  is generally caused by X-ray irradiation or collisions. 
  The electron transitions between the K-, L-shells and the outer shells 
  are the major processes in producing the keV emission lines. 
The strength of the lines are therefore determined 
  by the temperature, density, and the ionisation structures 
  of the emission region.   

The electron temperature at the accretion shock is $\sim 10^8$~K.  
This temperature is sufficient to completely ionise the elements 
  such as Argon, Silicon, Sulphur, Aluminium or Calcium 
  through electron collisions. 
Iron can be ionised to the highest ionisation states: 
  Fe~XXVI and Fe~XXVII. 

The ionisation due to irradiation can be described 
  by the ionisation parameter, 
  which is defined as $\xi = L_{\rm x}/n_{_{\rm H}} r^2$ 
  (Kallman \& McCray 1982). 
(Here, $L_{\rm x}$ is the luminosity of the irradiation, 
  $n_{_{\rm H}}$ is the number density of hydrogen, 
  and $r$ is the characteristic size of the plasma.) 
In the shock-heated region of mCVs, 
  $L_{\rm x} \sim 10^{31} - 10^{33}~{\rm erg}~{\rm s}^{-1}$, 
  $n_{_{\rm H}} \sim 10^{15} - 10^{16}~{\rm cm}^{-3}$, and 
  $r \sim 10^8~{\rm cm}$ and the ionisation parameter is  $\LS\ 10^2$. 
The corresponding ionisation states of Iron 
  are Fe~XX or lower (Makishima 1986). 

Therefore, in the shock-heated regions of mCVs  
  collisional ionisation is the major process 
  producing the keV emission lines. 
Photo-ionisation is important 
  only in the cooler and less dense pre-shock accretion stream.   
  
\subsection{Collisional ionisation equilibrium }   

For $T\sim 10^7 - 10^8$~K, 
  collisional ionisation equilibrium can be attained 
  when $n_{\rm e} t > 10^{12}\ {\rm cm}^{-3} {\rm s}$, 
  where $n_{\rm e}$ is the electron number density 
  and $t$ is the time spent by the ions in the plasma 
  (Gorenstein, Harnden \& Tucker 1974; Masai 1984). 
For a white dwarf with a specific accretion rate $\dot m$, 
  the local  value of $n_{\rm e} t$ at $\zeta$ 
  (the distance normalised to the shock  height $x_{\rm s}$) is  
\begin{eqnarray}   
  \big[ n_{\rm e} t \big]_{\zeta} & \approx  & 
      \bigg| {{\dot m} \over {m_{\rm H} v}} 
          \int^{v_{\rm ff}/4}_{v} {{dv'} \over v'} 
           {{dx} \over {dv'}} 
           \bigg|
           \nonumber  \\ 
     & = & {{x_{\rm s} R_{\rm w} \dot m } 
     \over {2 G M_{\rm w} m_{_{\rm H}}}} {1 \over \tau} 
   \int^{1/4}_\tau  {{d\tau'} \over \tau'} 
     \bigg( {{d\zeta} \over {d\tau'}} \bigg)\  , 
\end{eqnarray}  
  where $\tau = -v/v_{_{\rm ff}}$ is the dimensionless velocity, 
  and $v_{_{\rm ff}}$ is the free-fall velocity 
  at the white-dwarf surface.  

To evaluate the integral in the above equation 
  we consider the inverted velocity profile of the accretion flow 
  given in the hydrodynamic model of Wu (1994)   
  (see Cropper, Ramsay \& Wu 1998 for typographical corrections): 
\begin{equation}
 {{d\zeta}\over {d\tau}} = {2 v^2_{_{ff}}
    \over {x_{\rm s} A \rho_{\rm s}}} 
   {{\tau^2(5-8\tau)}\over{\sqrt{\tau(1-\tau)}}} K(\tau)\  ,     
\end{equation}  
  where $K(\tau) = 1+3^{-\alpha}4^{\alpha+\beta} 
        \epsilon_{\rm s} (1-\tau)^\alpha\tau^{\beta}$ 
  (with $\alpha \approx 2$ 
  and $\beta \approx 3.85$ appropriate for cyclotron cooling), 
  $\rho_{\rm s}$ is the density at the shock surface, 
  $\epsilon_{\rm s}$ is the ratio of the bremsstrahlung-cooling time-scale 
  to the cyclotron-cooling time-scale at the shock, 
  and $A = 3.9 \times 10^{16}$ in c.g.s.\ units. 
The shock-height $x_{\rm s}$ can be obtained 
  by a direct integration of the inverted velocity profile, i.e.,\ 
\begin{equation}
  x_{\rm s} = {2 v^2_{_{\rm ff}}\over {A \rho_{\rm s}}} 
     \int^{1/4}_0 d\tau 
      {{\tau^2(5-8\tau)}\over{\sqrt{\tau(1-\tau)}}} K(\tau) \  .      
\end{equation}  
In this formulation, the hydrodynamic variables, 
  such as density $\rho$, pressure $P$ and temperature $T$, 
  are proportional to $1/\tau$, $(1-\tau)$, and $\tau(1-\tau)$ respectively 
  (see Wu 1994; Wu, Chanmugam \& Shaviv 1994).   

By combining Equations (1) and (3), this yields 
\begin{eqnarray} 
\big[ n_{\rm e} t \big]_{\zeta}  
   \approx 2.2 \times 10^{16}\ F_1(\zeta;\epsilon_{\rm s})  \nonumber 
\end{eqnarray} 
\begin{eqnarray}
 \hspace{1cm}\times \bigg({M_{\rm w} \over {\rm M}_\odot}\bigg)^{1/2} 
    \bigg({R_{\rm w} \over {5 \times 10^8\ {\rm cm}}} \bigg)^{-1/2} 
  {\rm cm}^{-3} {\rm s} \ ,  
\end{eqnarray}  
where  
\begin{equation} 
     F_1(\zeta;\epsilon_{\rm s}) = {1 \over {\tau(\zeta)}} 
     \int^{1/4}_{\tau(\zeta)} d\tau'
        {{\tau'(5-8\tau')}\over{\sqrt{\tau'(1-\tau')}}} K(\tau')\ .    
\end{equation}  
As $t\sim x_{\rm s}/v_{_{\rm ff}} 
   \sim n_{\rm e}^{-1}~{\rm min}(1,\epsilon_{\rm s}^{-1})$,     
when $\epsilon_{\rm s}$ is specified  
  $[n_{\rm e} t]_{\zeta}$ does not explicitly depend on $\dot m$ 
  (see Equation 4). 
In Figure~1, we show $2.2\times 10^4\ F_1(\zeta;\epsilon_{\rm s})$ 
  as a function of $(1-\zeta)$, 
  the distance from the shock surface. 
When $2.2 \times 10^4\ F_1(\zeta;\epsilon_{\rm s}) > 1$, 
  the accreting matter is in collisional ionisation equilibrium. 
Figure 1 shows that except in a very thin layer 
  near the shock front itself,  
  collisional ionisation equilibrium is reached  
  in the post-shock region 
  for the range of $\epsilon_{\rm s}$ of interest.  

\begin{figure}
\begin{center}        
\epsfxsize=8cm  
\epsfbox{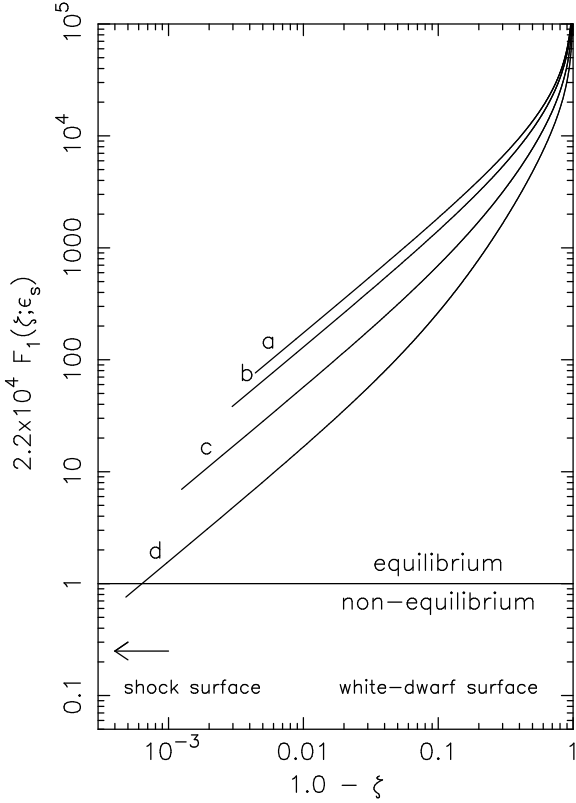}   
\end{center}        
\caption{The quantity $2.2 \times 10^4 F_1(\zeta;\epsilon_{\rm s})$ 
   as a function of $(1-\zeta)$, 
   the normalised distance from the shock surface.  
   The plasma is in collisional equilibrium   
   if $2.2 \times 10^4 F_1(\zeta;\epsilon_{\rm s}) > 1$.   
   Curves a, b, c and d correspond to 
   $\epsilon_{\rm s} =$ 0, 1, 10 and 100 respectively.   } 
\label{fig.1 }             
\end{figure}            
   
\subsection{Ionisation balance}  

\begin{figure} 
\begin{center}        
\epsfxsize=8cm  
\epsfbox{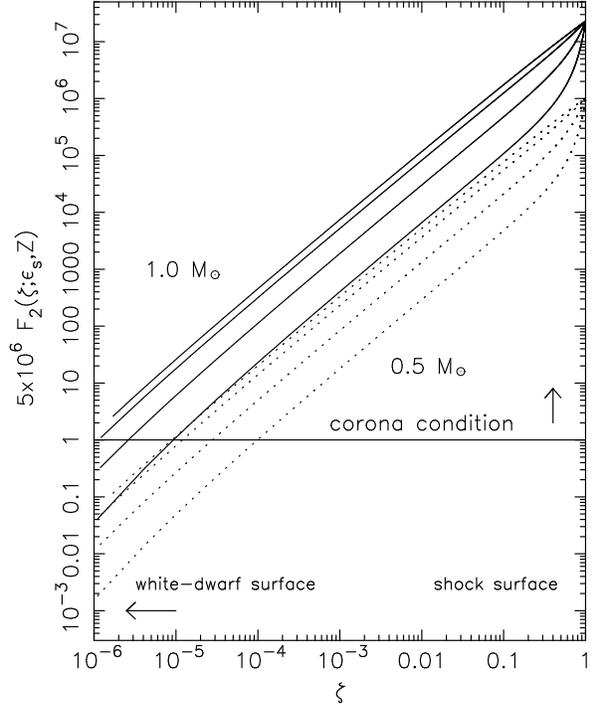}   
\end{center}        
\caption{$5 \times 10^6\ F_2(\zeta;\epsilon_{\rm s},Z)$ 
    as a function of $\zeta$, 
    the normalised height from the white-dwarf surface, 
    for Iron ($Z=26$). 
  For the solid curves $M_{\rm w} = 1.0 {\rm M}_\odot$ and 
    $R_{\rm w} = 5.53 \times 10^8~{\rm cm}$; 
    and for the dotted curves, $M_{\rm w} = 0.5 {\rm M}_\odot$ and 
    $R_{\rm w} = 9.61 \times 10^8~{\rm cm}$. 
  In each set of the curves, 
    the corresponding $\epsilon_{\rm s}$ are 0, 1, 10 and 100 
    (from top to bottom respectively). 
    The horizontal line represents 
    $F_2(\zeta;\epsilon_{\rm s},Z) = 2.0 \times 10^{-7}$, 
    above which the approximation of corona condition is applicable.  } 
\label{fig.2 }       
\end{figure}    

The ionisation balance is determined 
  by the excitation and de-excitation rates 
  of the corresponding ionisation states. 
As the rate coefficients may depend on the optical depth, 
  it is generally required to solve 
  the coupled population rate and radiative-transfer equations 
  to obtain the population of the ionisation states 
  (Bates, Kingston \& McWhirter 1962; Castor 1993). 
However, if the electron number density is sufficiently low 
  that the radiative-transfer effects are unimportant, 
  the ionisation balance can be calculated 
  using the corona-condition approximation. 
In the other extreme, 
  if the electron number density is sufficiently high 
  such that the plasma is able 
  to attain local thermal equilibrium (LTE), 
  the population can then be determined by the Saha equation.   

For the corona condition to hold, 
  it requires the gas to be in collisional equilibrium  
  and a direct coupling of the heat input 
  to the ions and the free electrons. 
It also requires the density of the plasma 
  to be sufficiently low such that most ions 
  are in ground states, 
  and the line (and continuum) emission to be optically thin 
  (Elwert 1952; Mewe 1990; Kahn \& Liedahl 1991). 

For ions with a bared nuclear charge $Z$, 
  the corona condition is satisfied 
  if $n_{\rm e}\ \LS \ 4 \times 10^4 Z^2\ T_{\rm e}^2\ {\rm cm}^{-3}$ 
  (Wilson 1962; Mewe 1990). 
As $n_{\rm e}$ and $T_{\rm e}$ scale with $1/\tau$ and $\tau(1-\tau)$, 
  where $\tau$ is the dimensionless velocity (see \S 2.1), 
  we can define a quantity 
\begin{eqnarray}  
   F_2(\zeta;\epsilon_{\rm s}; Z) 
    = Z^2 \tau(\zeta)^3 [1 - \tau(\zeta)]^2 
    \bigg({\mu \over 0.5} \bigg)^2 
    \bigg({{\dot m}\over {1\ {\rm g}\ {\rm s}^{-1}}} \bigg)^{-1}  
 \nonumber  
\end{eqnarray}
\begin{eqnarray}
  \hspace{1cm} \times \ \bigg({M_{\rm w} \over {\rm M}_\odot} \bigg)^{5/2}
    \bigg({R_{\rm w} \over {5\times 10^8\ {\rm cm}}} \bigg)^{-5/2} \ .  
\end{eqnarray}   
The corona-condition criterion is then equivalent to require 
  $F_2(\zeta;\epsilon_{\rm s}; Z)~\GS\  2\times 10^{-7}$.   

In Figure 2 we show the profile 
  of the function $[5\times 10^6 F_2(\zeta; \epsilon_{\rm s},Z)]$ 
  in the shock-heated regions of accreting magnetic white dwarfs 
  for various parameters. 
It can be seen that for white dwarfs 
  with $M_{\rm w} = 1~{\rm M}_\odot$ 
  and $\dot m = 1~{\rm g}~{\rm s}^{-1}$, 
  the corona condition is satisfied 
  when the height $\zeta > 10^{-5}$.  
For white dwarfs with $M_{\rm w} = 0.5~{\rm M}_\odot$, $\zeta > 10^{-4}$. 
The corona-condition approximation is not applicable 
  at the very bottom of the shock-heated region, 
  where the electron number density is high 
  and the temperature is low. 

A plasma is in LTE 
  when $n_{\rm e}~\GS~1.4 \times 10^{15} Z^6 T^{1/2}~{\rm cm}^{-3}$ 
  (Wilson 1962). 
In terms of the velocity profile 
  of the post-shock accretion flow and white-dwarf parameters, 
  it requires    
\begin{eqnarray}  
    1\ \GS\ 9.7\times 10^4  Z^6 \tau(\zeta)^{3/2} 
      [1 - \tau(\zeta)]^{1/2}
   \bigg({\mu \over 0.5} \bigg)^{1/2} \nonumber 
\end{eqnarray} 
\begin{eqnarray}
  \hspace{1cm}  \times\       
    \bigg({{\dot m} \over {1~{\rm g}~{\rm s}^{-1}}} \bigg)^{-1}
    \bigg( {M_{\rm w} \over {\rm M}_\odot} \bigg) 
    \bigg({R_{\rm w} \over {5\times 10^8\ {\rm cm}}} \bigg)^{-1}.    
\end{eqnarray}  
For Iron, $Z = 26$. LTE implies $\tau\ \LS\ 10^{-9}$, 
  corresponding to $v\ \LS\ 1~{\rm cm}~{\rm s}^{-1}$. 
Thus, LTE is applicable for the very base of the post-shock flow only.  

\begin{figure} 
\centering 
\epsfxsize=8.2cm       
\epsfbox{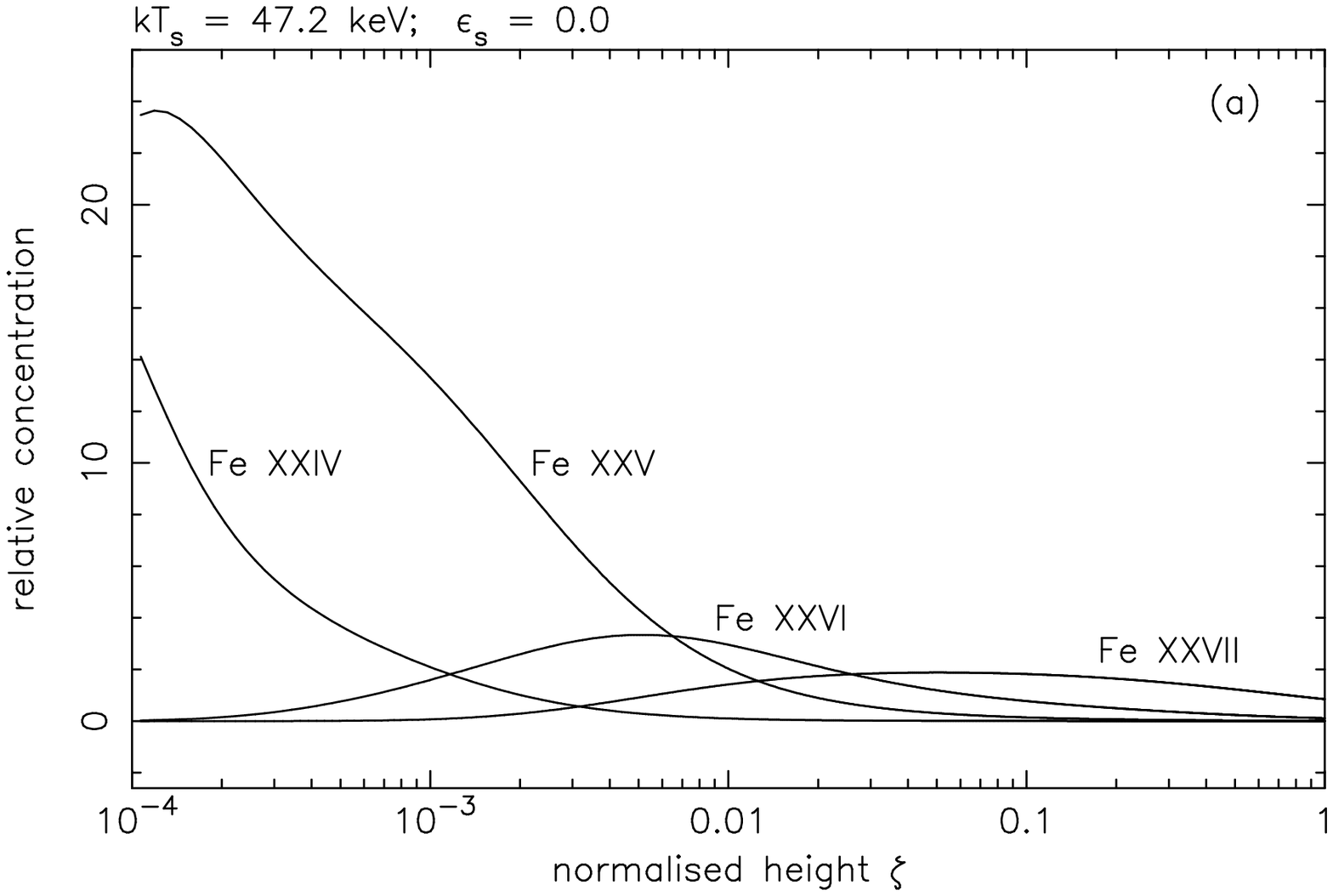}  
\epsfxsize=8.2cm    
\epsfbox{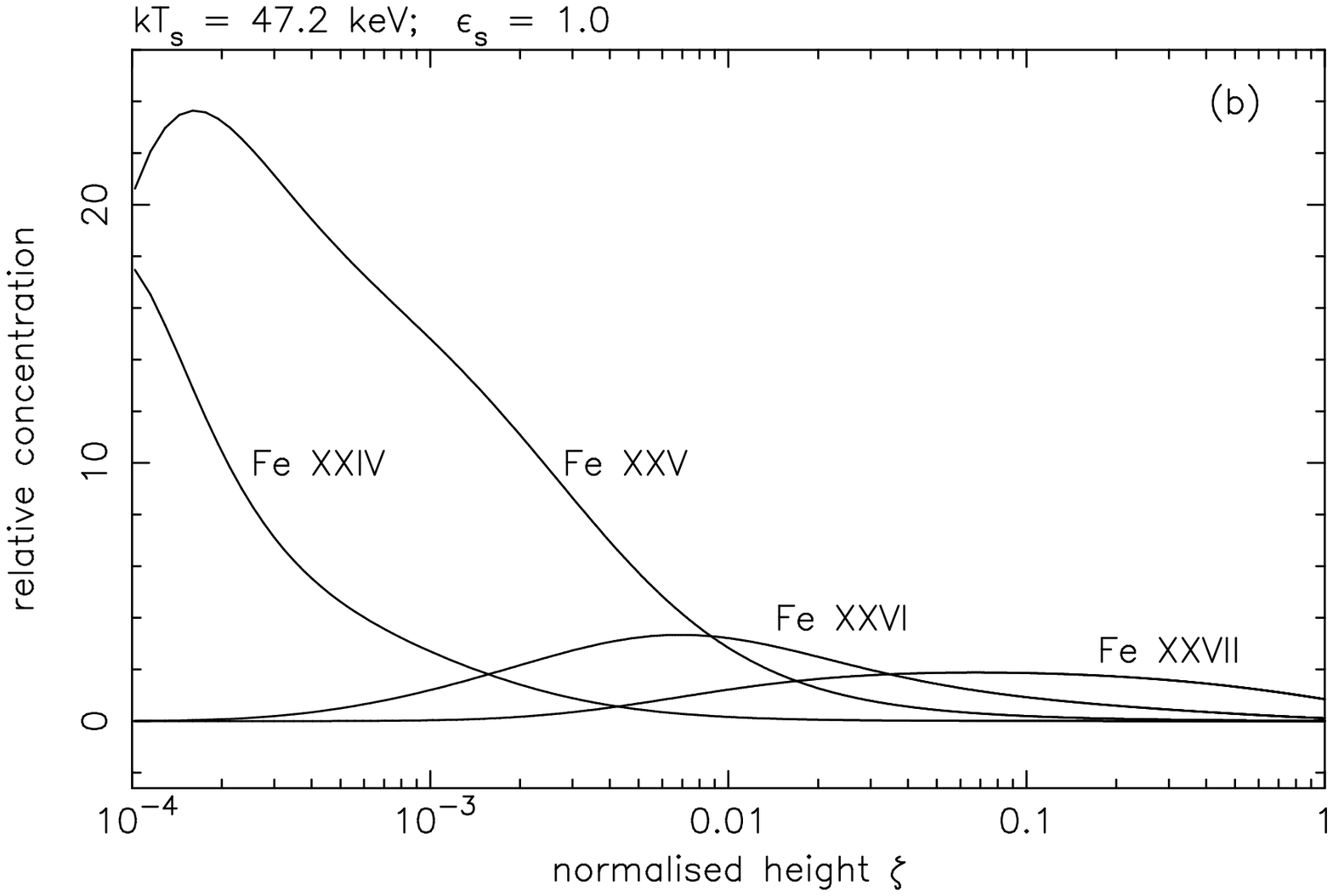}  
\epsfxsize=8.2cm       
\epsfbox{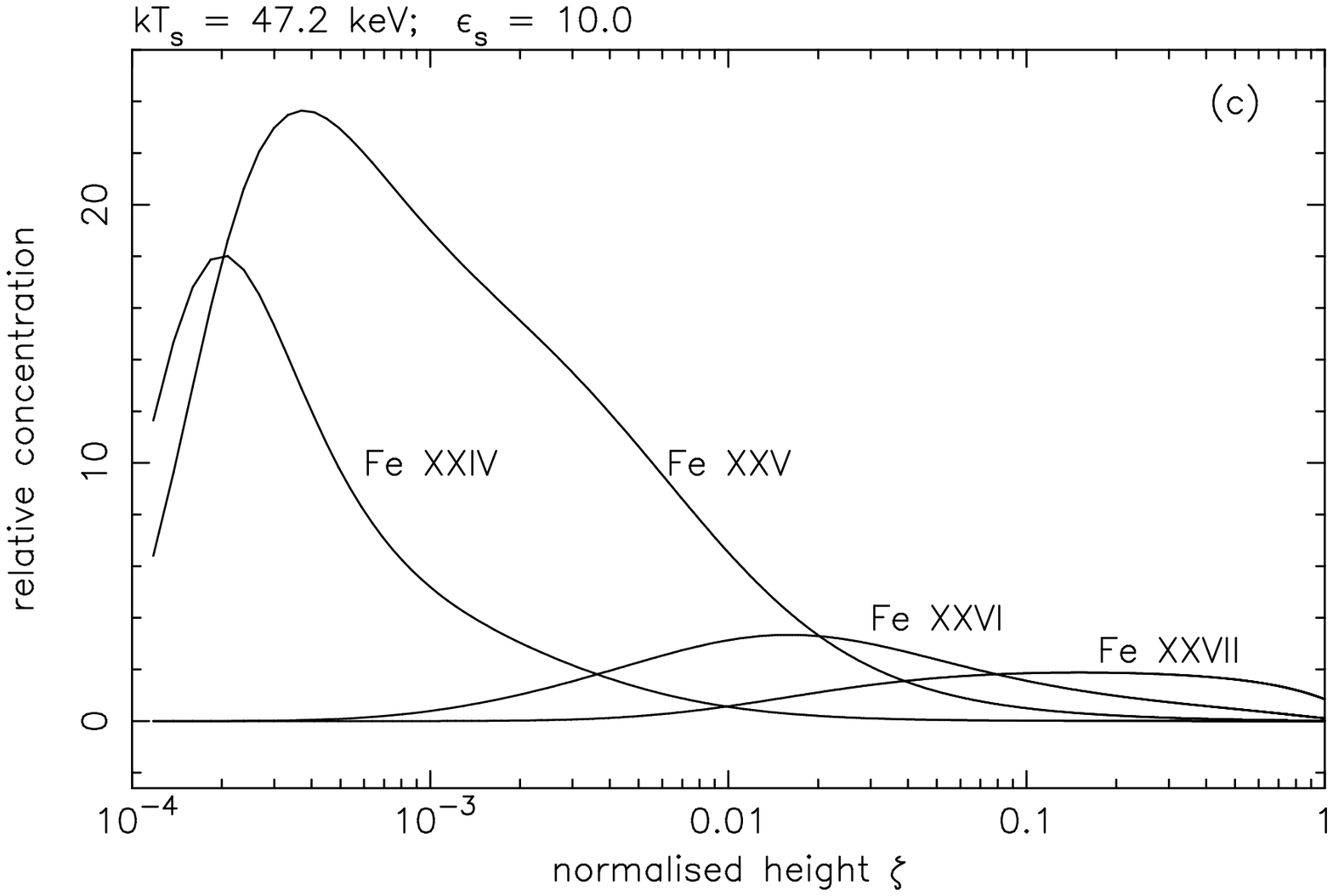}  
\epsfxsize=8.2cm    
\epsfbox{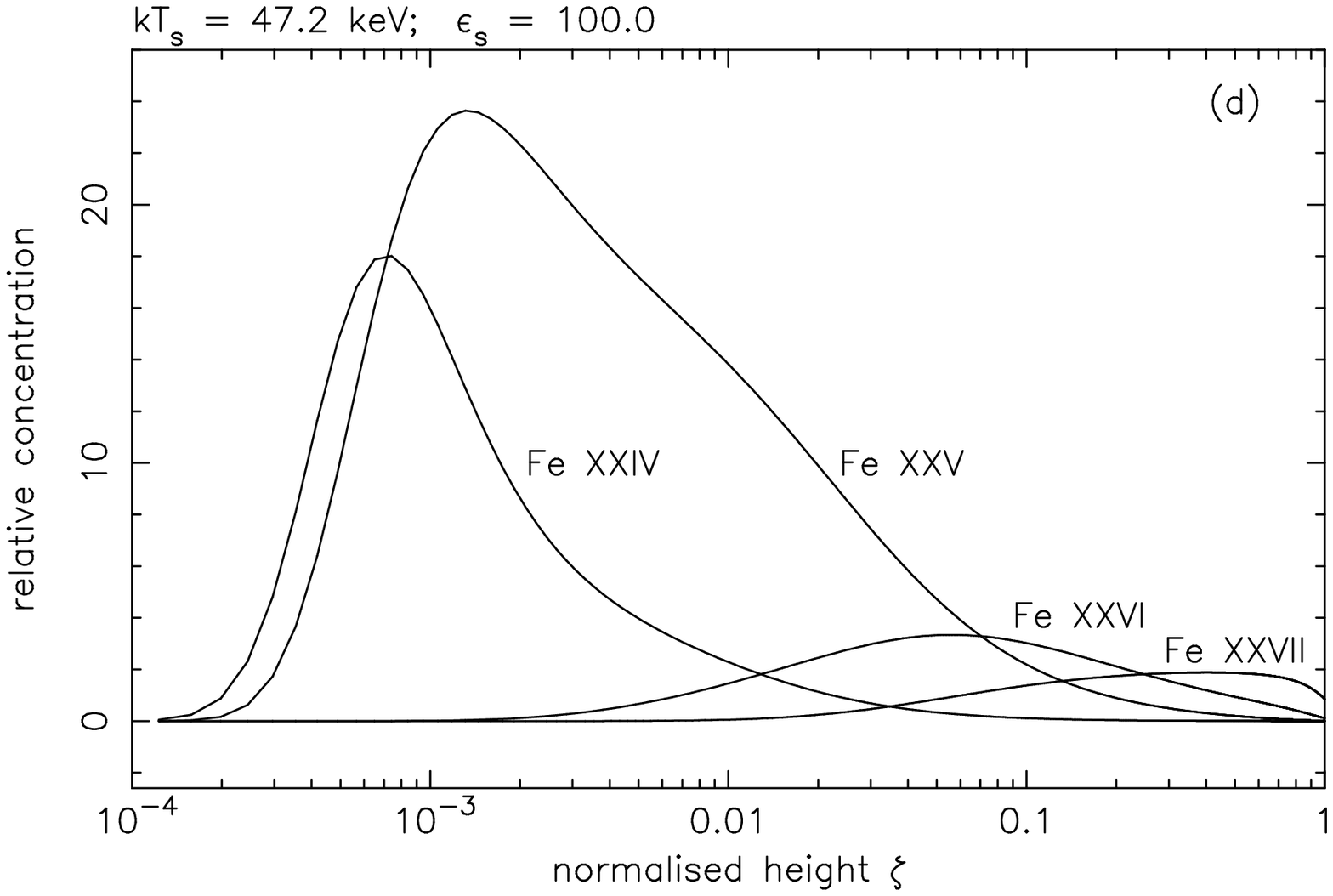}           
\caption{The ionisation profile of Fe~XXVII, Fe~XXVI, Fe~XXV and Fe~XXIV 
    in the shock-heated emission region 
    of a 1.0-M$_\odot$ white dwarf. 
  The effects of cyclotron cooling increases 
    from the top to the bottom panels 
    (from panel (a) to panel (d) respectively), 
    and the corresponding parameters 
    are $\epsilon_{\rm s} = $ 0, 1, 10  and 100. 
  The surface of the white dwarf is at $\zeta = 0$, 
    and the shock surface is at $\zeta =1$.   } 
\label{fig.3 }       
\end{figure}    

\begin{figure} 
\centering 
\epsfxsize=8.2cm       
\epsfbox{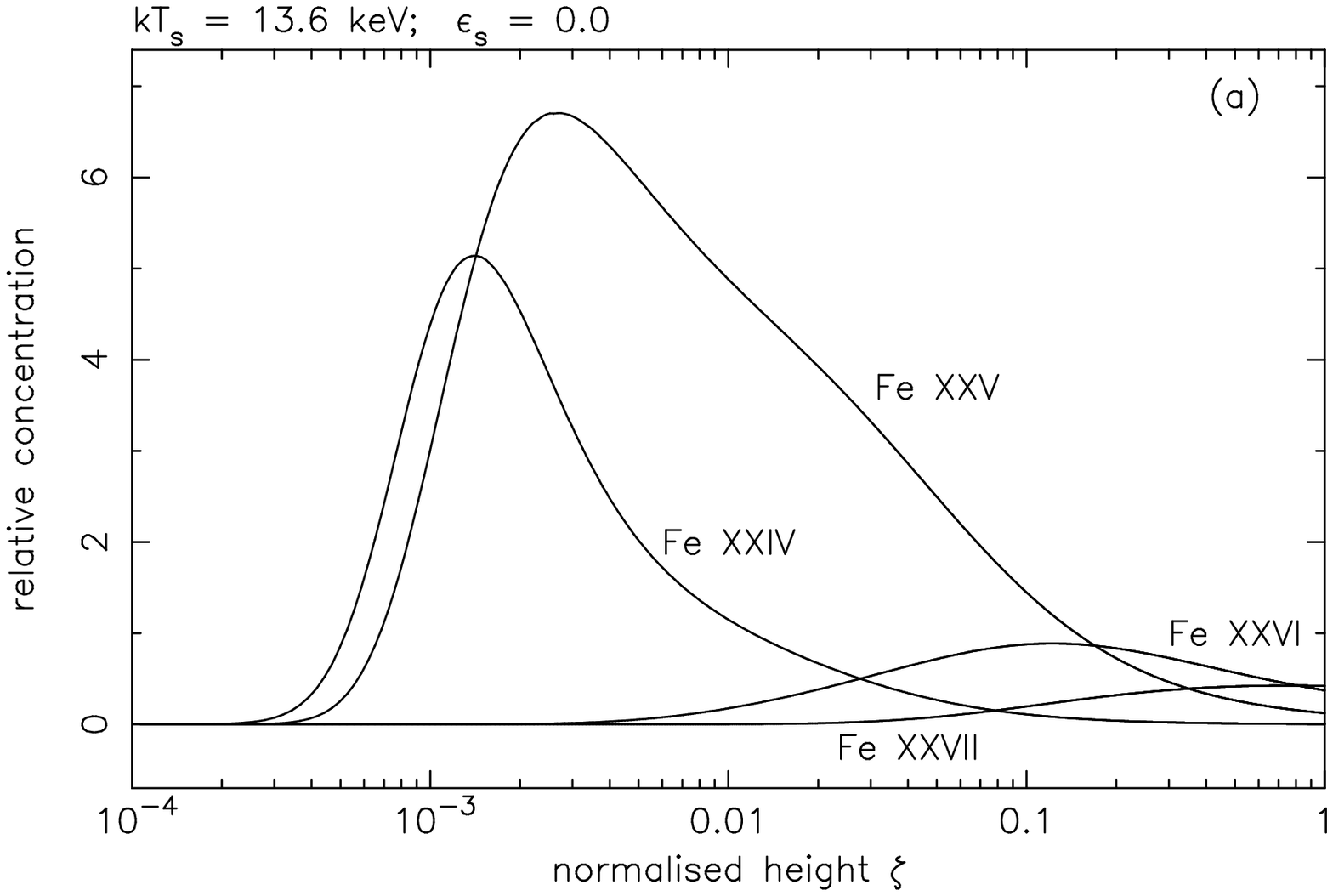}  
\epsfxsize=8.2cm    
\epsfbox{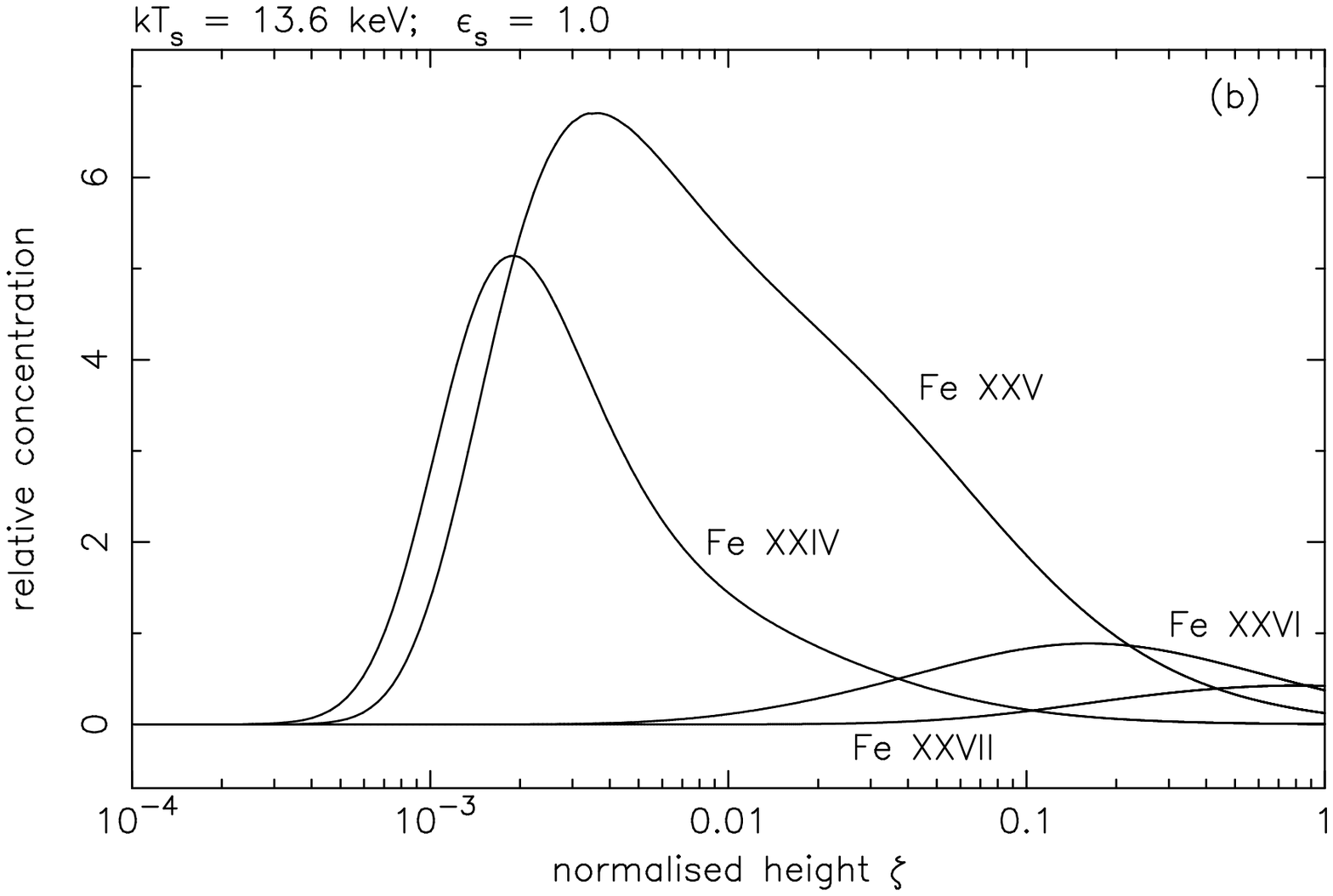}  
\epsfxsize=8.2cm       
\epsfbox{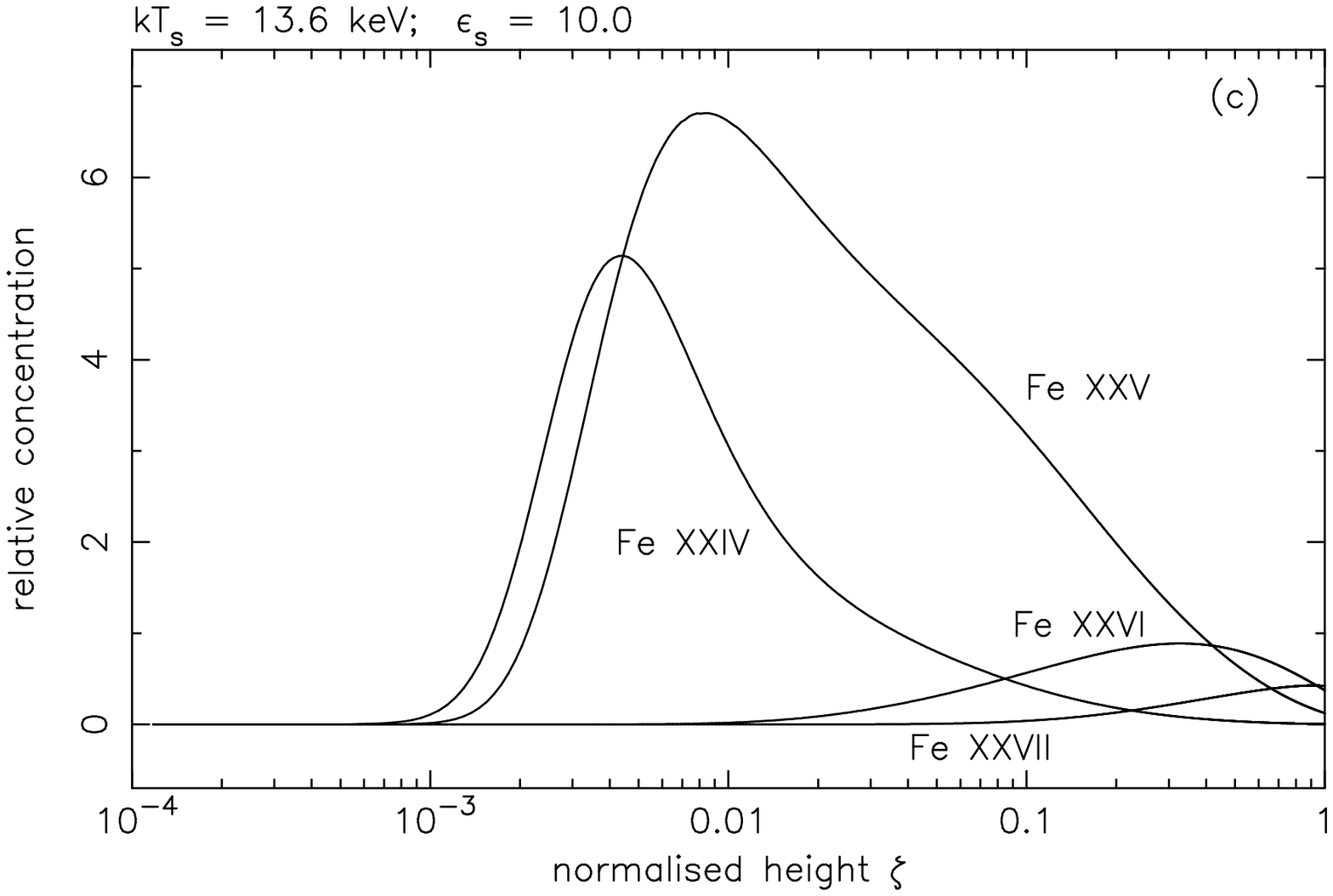}  
\epsfxsize=8.2cm    
\epsfbox{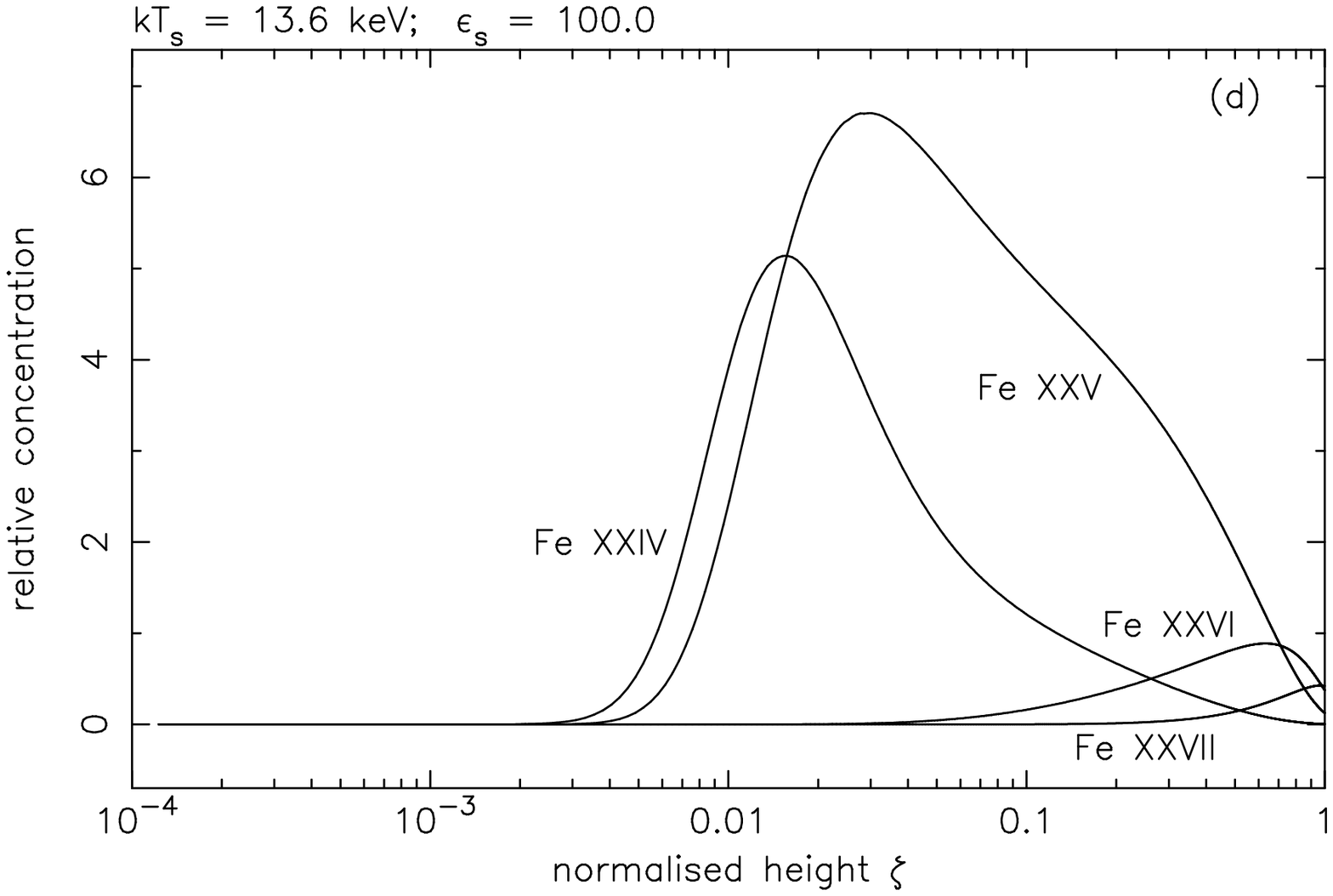}           
\caption{Same as Fig.~3 for a white dwarf with 0.5~M$_\odot$.  } 
\label{fig.4 }       
\end{figure}   

\begin{figure} 
\centering 
\epsfxsize=8.2cm       
\epsfbox{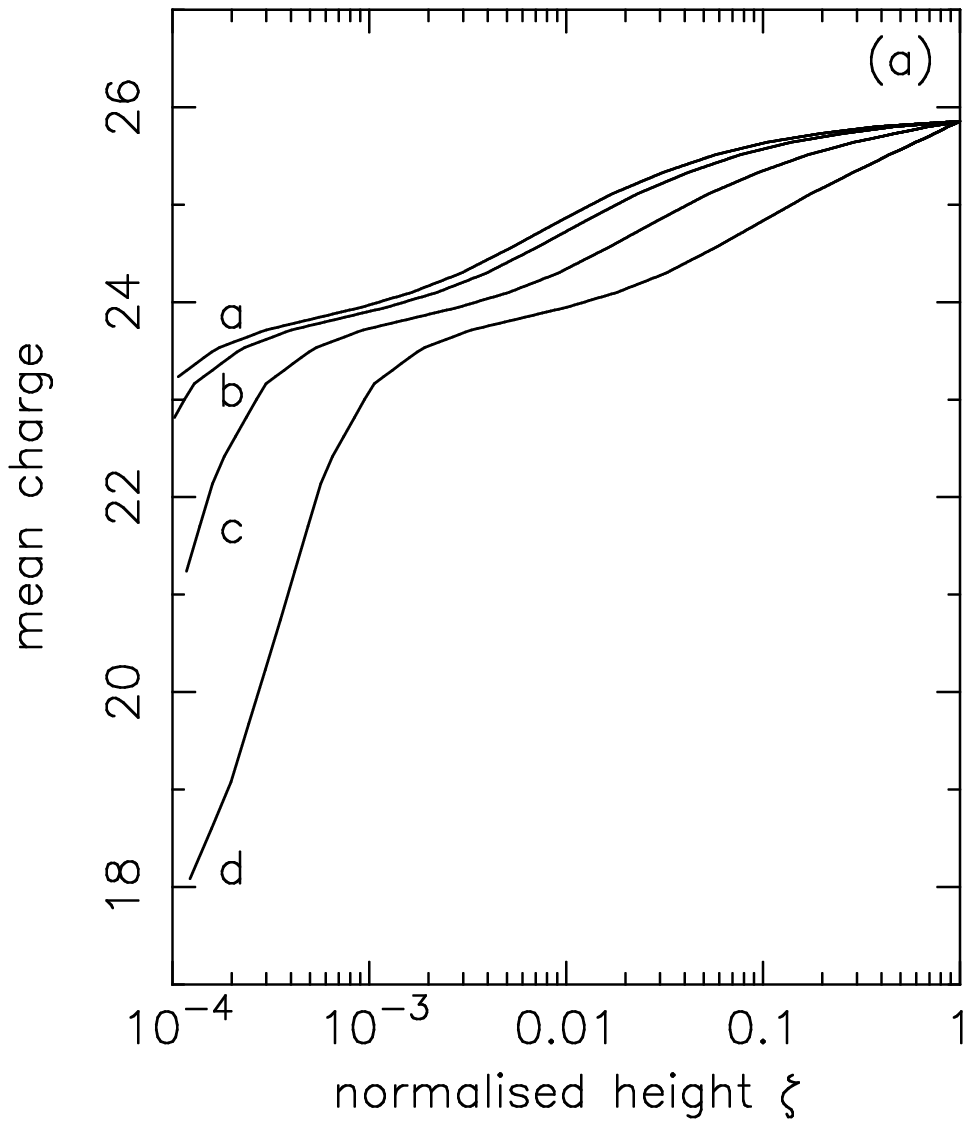}  
\epsfxsize=8.2cm    
\epsfbox{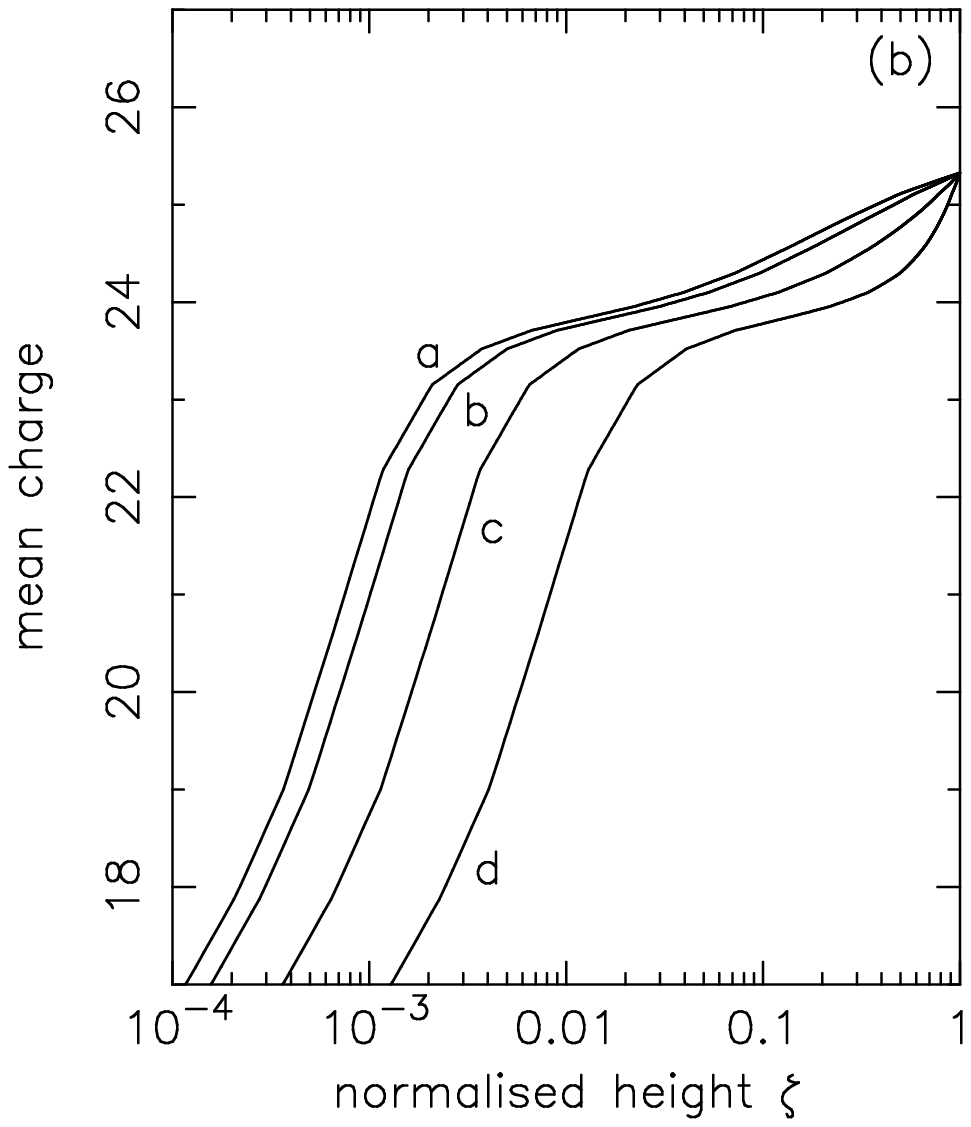}           
\caption{he mean charge profile of Iron in the shock-heated region 
    for (a) $M_{\rm w}  = 1.0 {\rm M}_\odot$; 
    and (b) for $M_{\rm w}= 0.5 {\rm M}_\odot$. 
  Curves a, b, c and d correspond 
    to $\epsilon_{\rm s} = $ 0, 1, 10 and 100 respectively.  }   
\label{fig.5 }  
\end{figure}     
 
\subsection{Ionisation profiles}    

Ionisation-balance calculations in the approximation of corona condition   
  have been carried out by many authors (e.g.\ Jordan 1969;  
  Jacobs et al.\ 1977; Raymond \& Smith 1977; Mewe \& Gronenschild 1981;  
  Arnaud \& Rothenflug 1985; Mewe, Gronenschild \& van den Oord 1985). 
The results of the calculations are usually presented   
  in terms of $n_{\rm z}/\sum_{\rm z} n_{\rm z}$ 
  (the relative ion concentration) 
  as a function of $T_{\rm e}$ (the electron temperature). 
With the local value for the electron number density, 
  the electron temperature and the abundance of the elements specified, 
  the results of these ionisation balance calculations 
  can be readily used to determine the concentration of the ion species.   

The temperature of the accretion shock in mCVs  
  is typically $\sim 10^8$~K, which is sufficient to ionise Iron 
  in the shock-heated region to its higher ionisation states.  
As only the most populous ion species 
  would make significant contribution to the  emission, 
  in this study we only consider the four most highly ionisation states, 
  Fe~XXVII (Fe$^{26+}$), Fe~XXVI (Fe$^{25+}$), 
  Fe~XXV (Fe$^{24+}$) and Fe~XXIV (Fe$^{23+}$). 
Also, we restrict the emission region of interest 
  to be $\zeta > 10^{-4}$, where the corona condition is satisfied.   

We first determine the local ionisation 
  by interpolating the results  
  from  the ionisation balance calculations of Arnaud \& Rothenflug (1985)  
  and then use it to calculate the profiles 
  of the relative concentration of the ion species 
  in the shock-heated region. 
Two representative cases are considered, 
  with the white-dwarf masses 1.0 and 0.5~M$_\odot$ respectively 
  for each case.  
(We have assumed the Nauenberg (1972) mass-radius in the calculations.) 
The parameter $\epsilon_{\rm s}$ is chosen to be 0, 1, 10 and 100. 
This covers the range of parameters 
  appropriate for most mCVs (see Wu, Chanmugam \& Shaviv 1994).

\begin{figure} 
\centering 
\epsfxsize=8.2cm       
\epsfbox{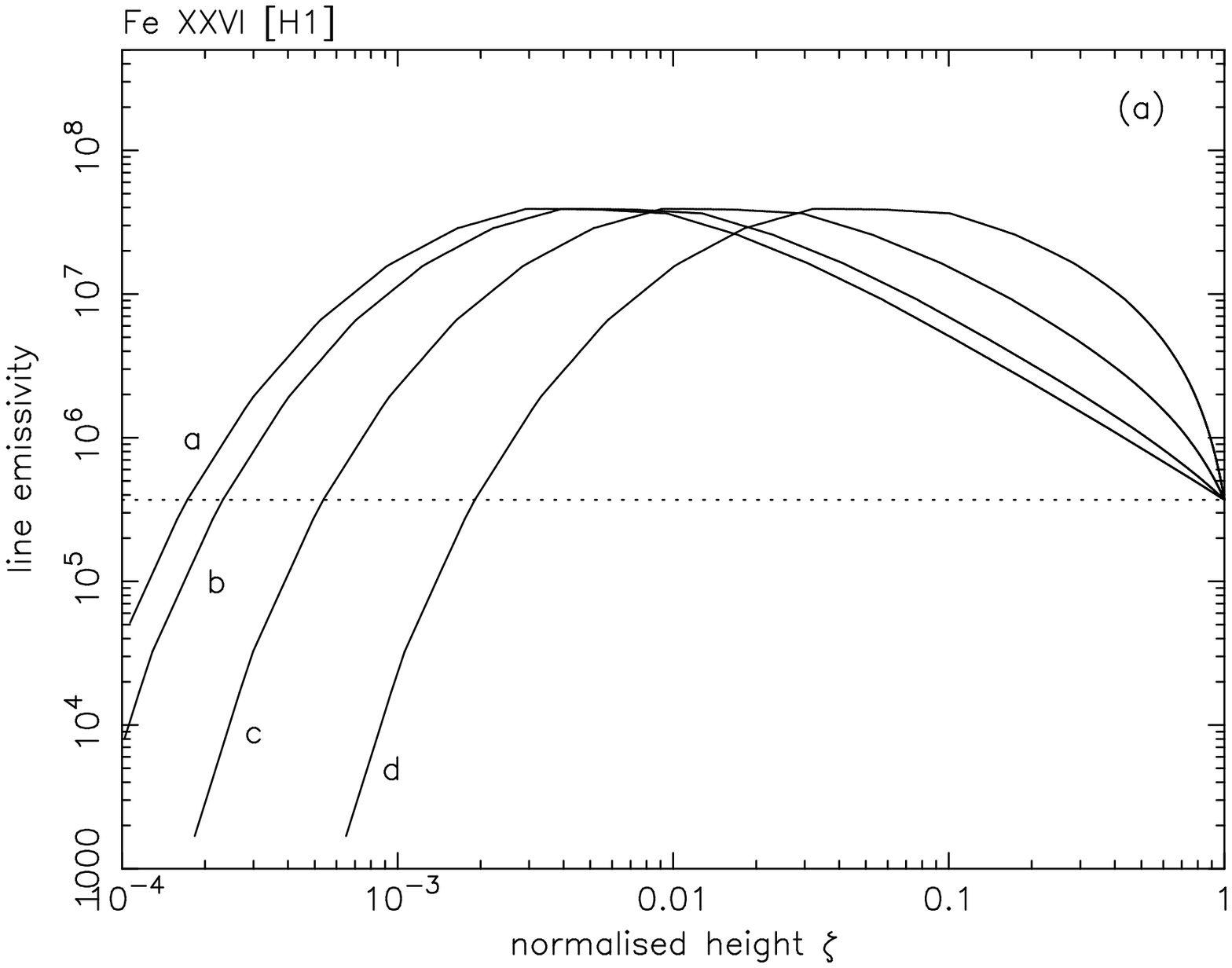}  
\epsfxsize=8.2cm    
\epsfbox{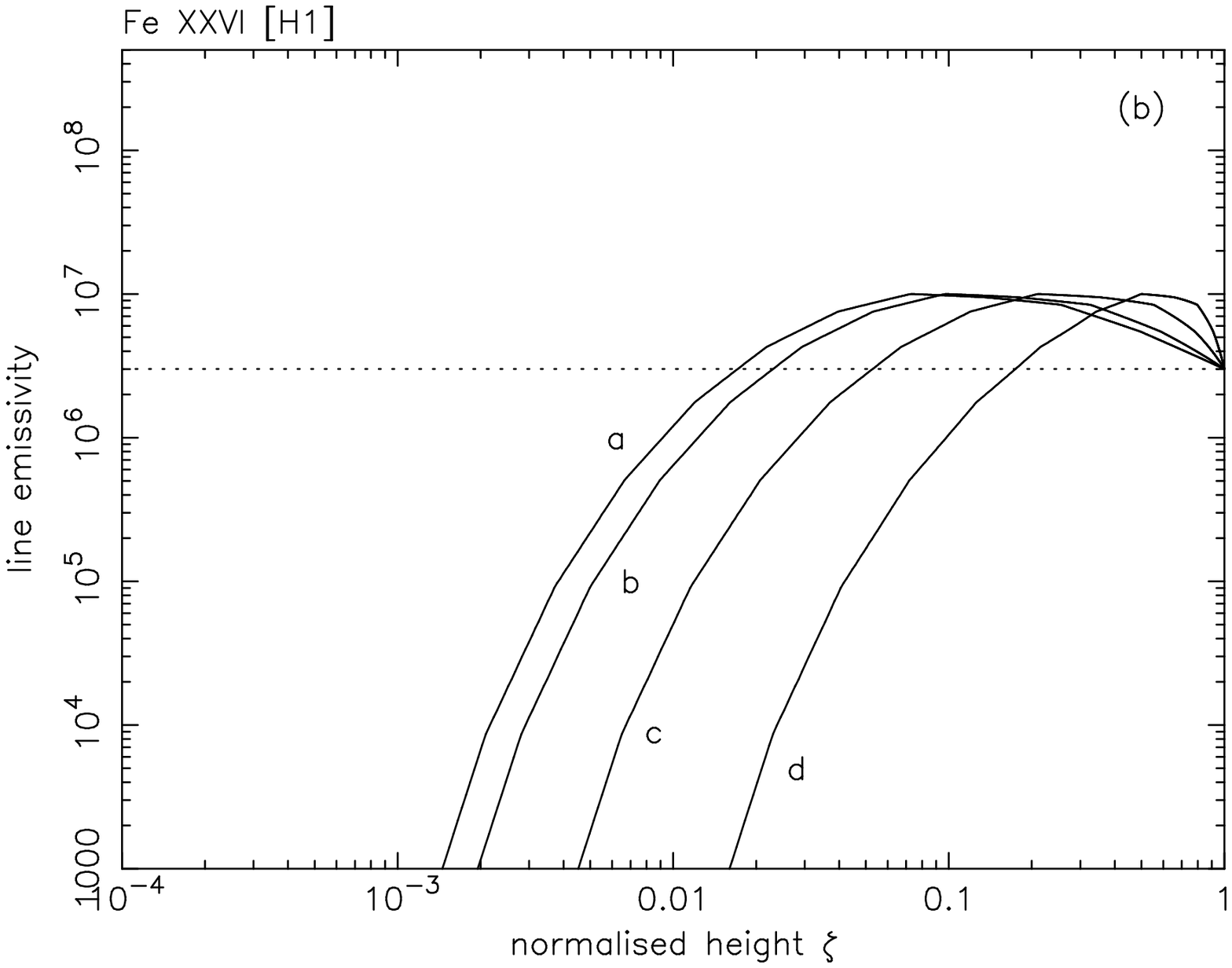}          
\caption{The emissivity (in erg~cm$^{-3}$~s$^{-1}$) profile 
    of the Fe~XXVI Lyman-$\alpha$ (1s $^2$S -- 2p $^2$P) transition 
    for a white dwarf with a mass of 1.0~M$_\odot$ (upper panel) and 
    0.5~M$_\odot$ (lower panel). 
  The specific accretion rate is 
    $\dot m = $ 1~g~cm$^{-2}$~s$^{-1}$. 
  Curves a, b, c and d correspond 
    to $\epsilon_{\rm s} =$ 0, 1, 10 and 100 respectively.  
  For comparison, we also show in dotted line 
    the line emissivity of emission regions      
    with constant electron temperature and number density 
    the same as the values at the shock 
    of the structured emission regions that we consider. } 
\label{fig.6 }     
\end{figure}    

\begin{figure} 
\centering 
\epsfxsize=8.2cm       
\epsfbox{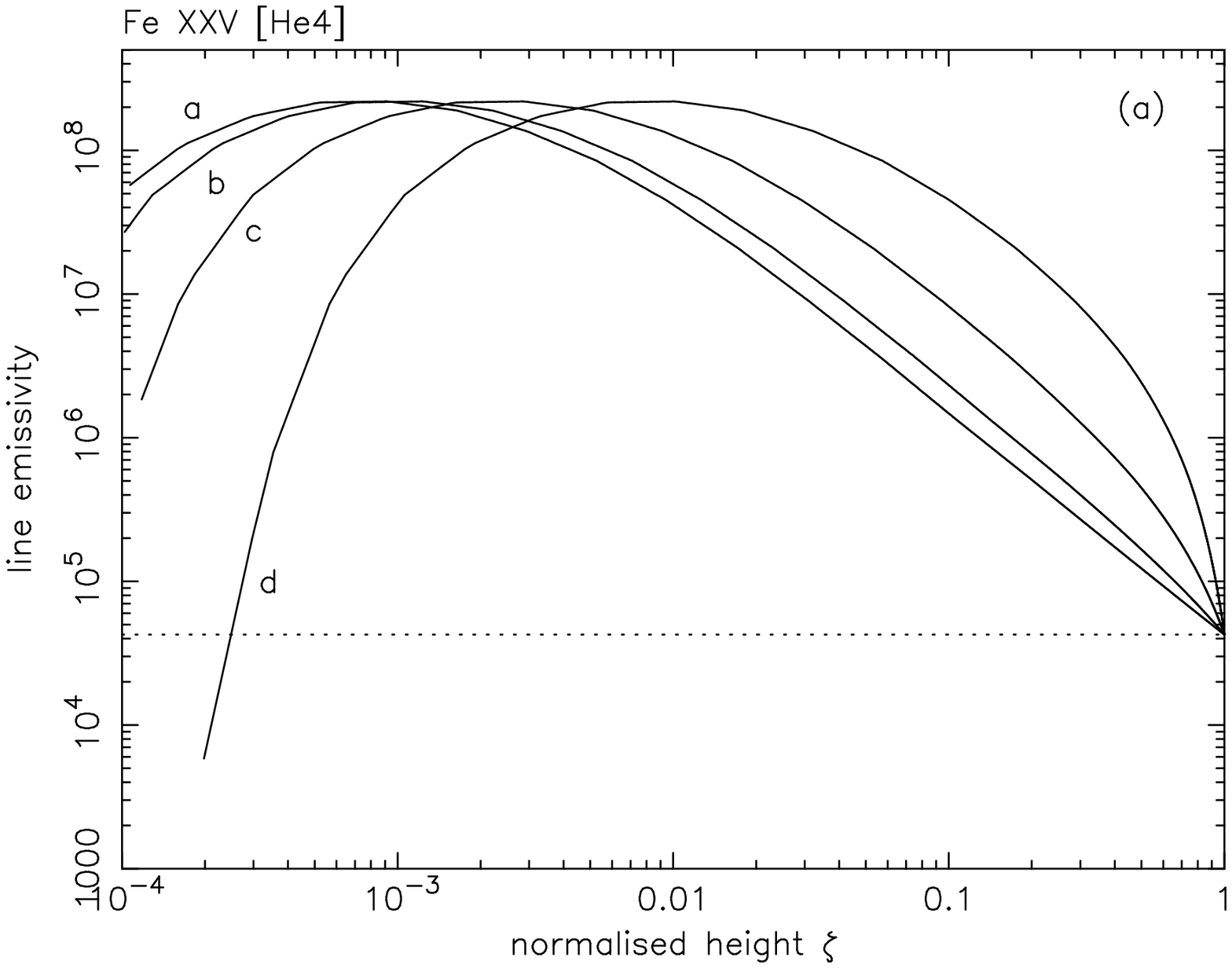}  
\epsfxsize=8.2cm    
\epsfbox{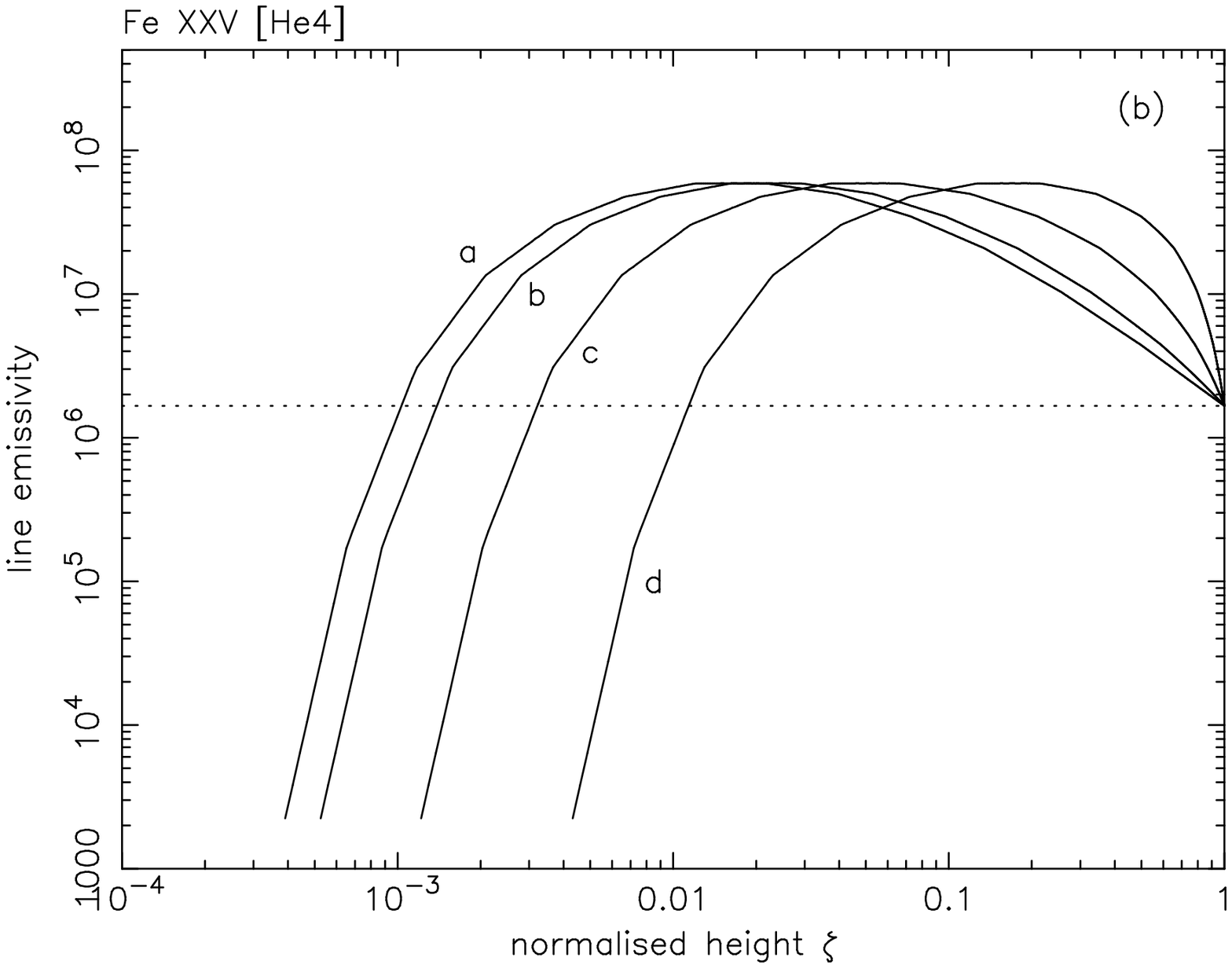}          
\caption{
   Same as Fig.~6 
     for the Fe~XXV He4 (1s$^2$ $^1$S -- 1s2p $^1$P) transition.  } 
\label{fig.7 }     
\end{figure}    

\begin{figure} 
\centering 
\epsfxsize=8.2cm       
\epsfbox{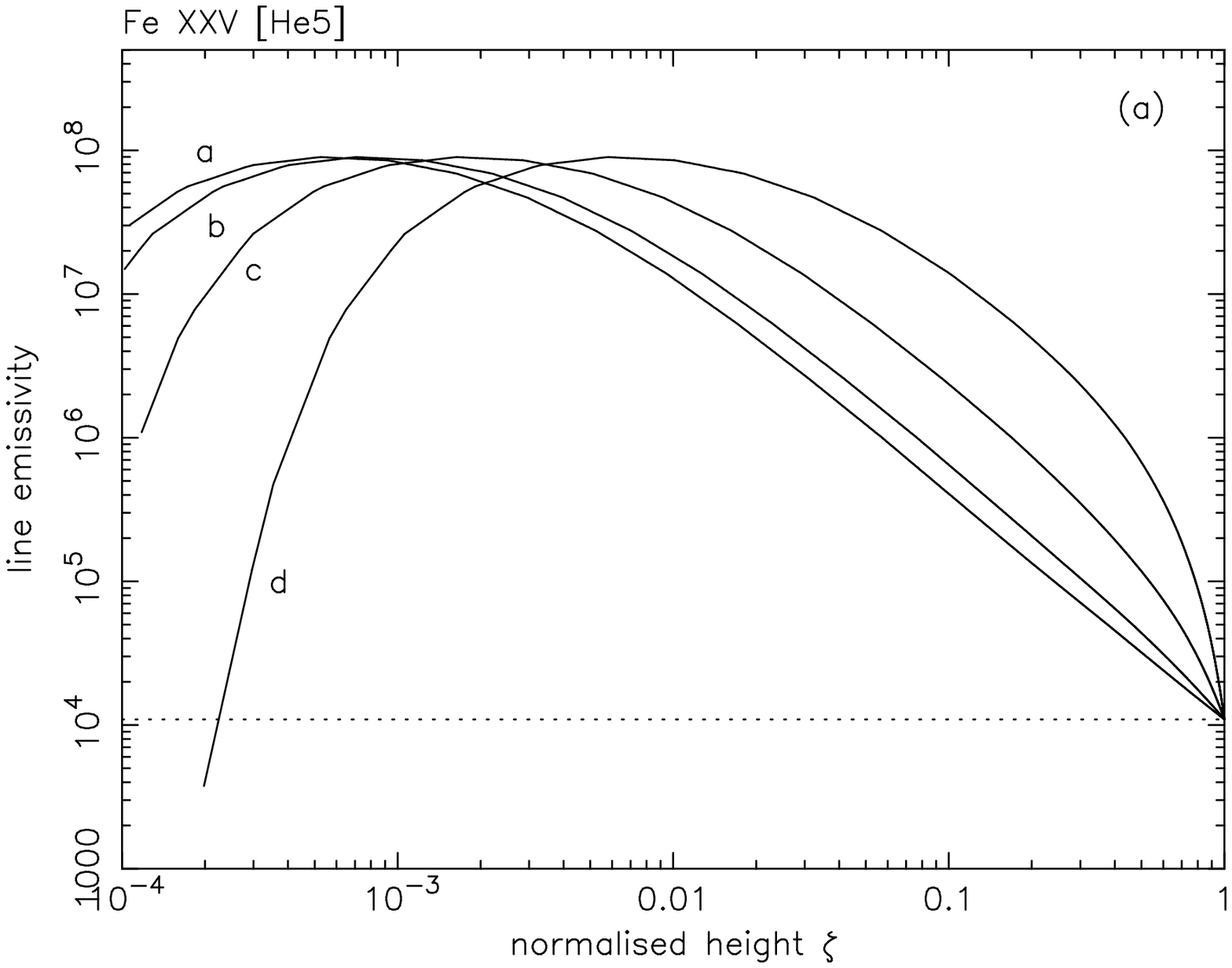}  
\epsfxsize=8.2cm    
\epsfbox{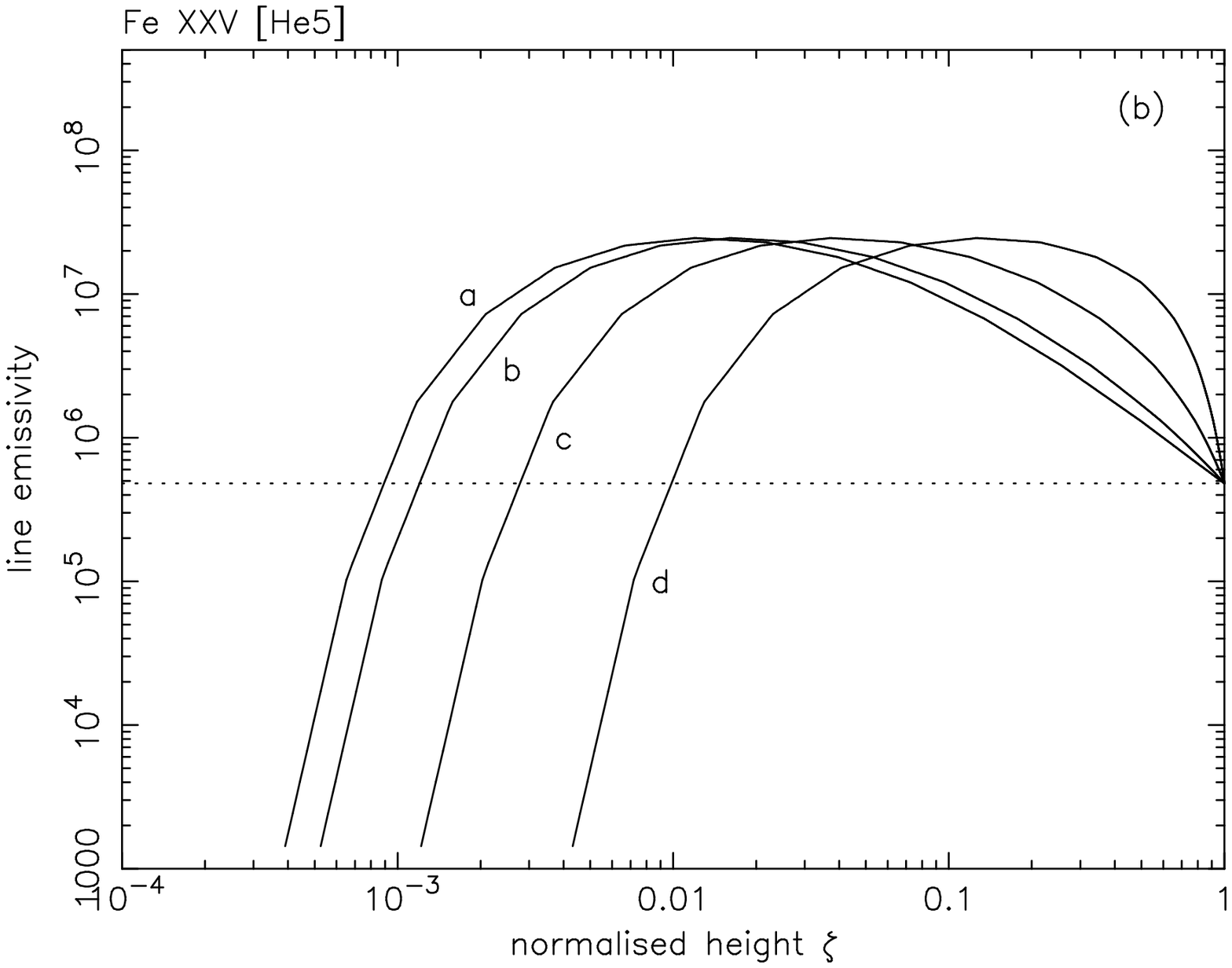}          
\caption{
   Same as Fig.~6 
      for the Fe~XXV He5 (1s$^2$ $^1$S -- 1s2p $^3$P) transition.  } 
\label{fig.8 }     
\end{figure}    

\begin{figure} 
\centering 
\epsfxsize=8.2cm       
\epsfbox{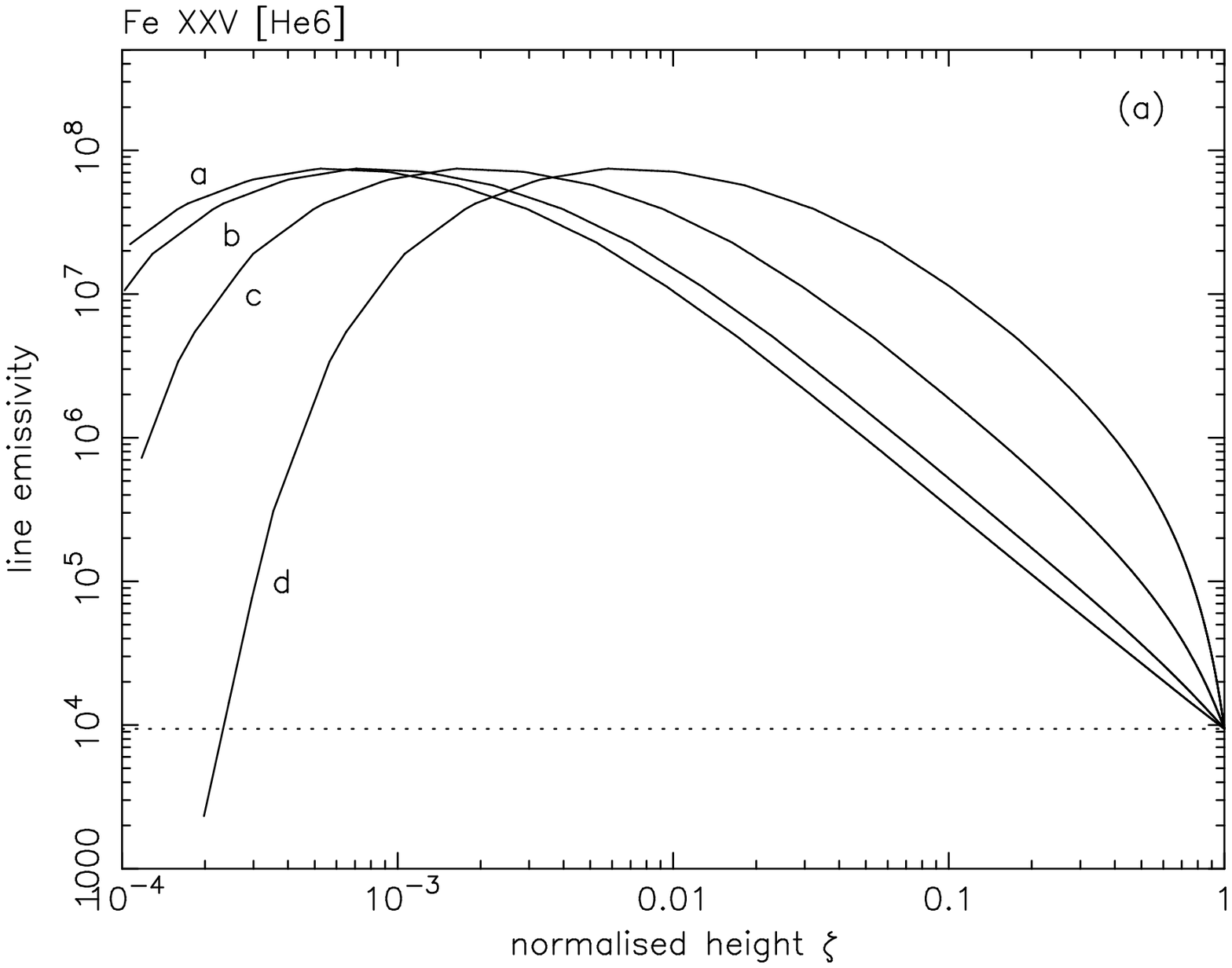}  
\epsfxsize=8.2cm    
\epsfbox{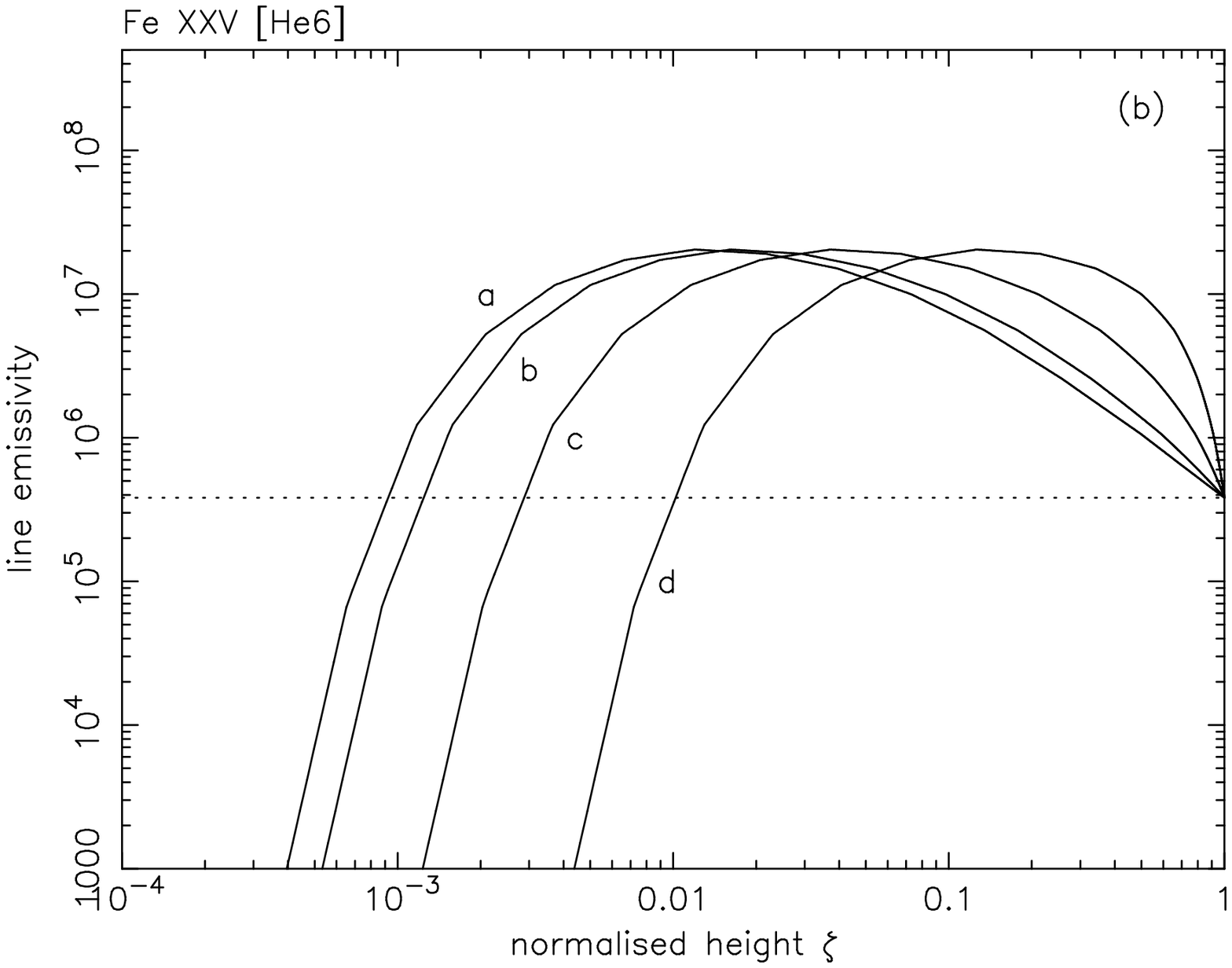}          
\caption{Same as Fig.~6 
  for the Fe~XXV He6 (1s$^2$ $^1$S -- 1s2p $^3$S) transition. } 
\label{fig.9 }     
\end{figure}   

\begin{figure} 
\centering 
\epsfxsize=8.2cm       
\epsfbox{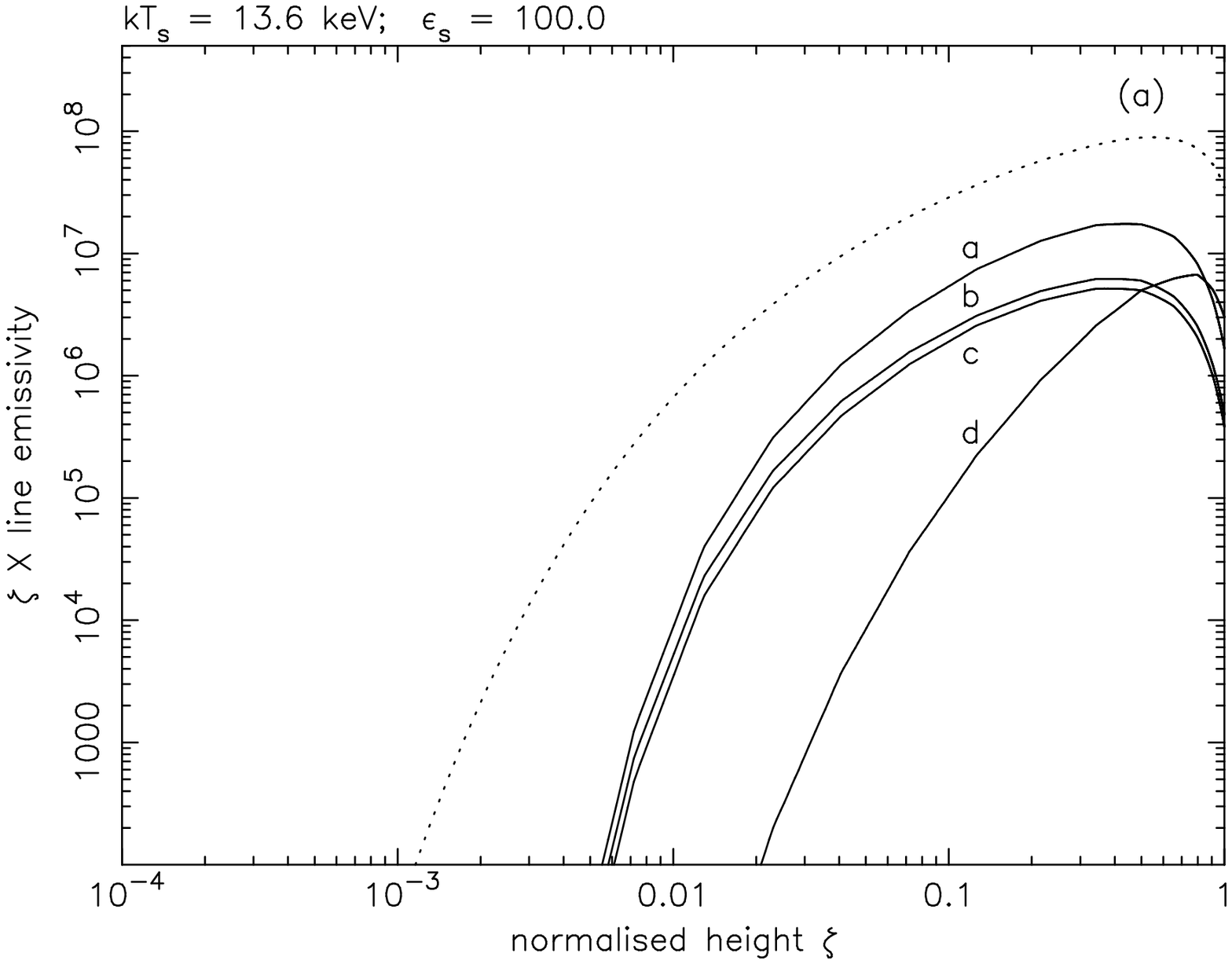}  
\epsfxsize=8.2cm    
\epsfbox{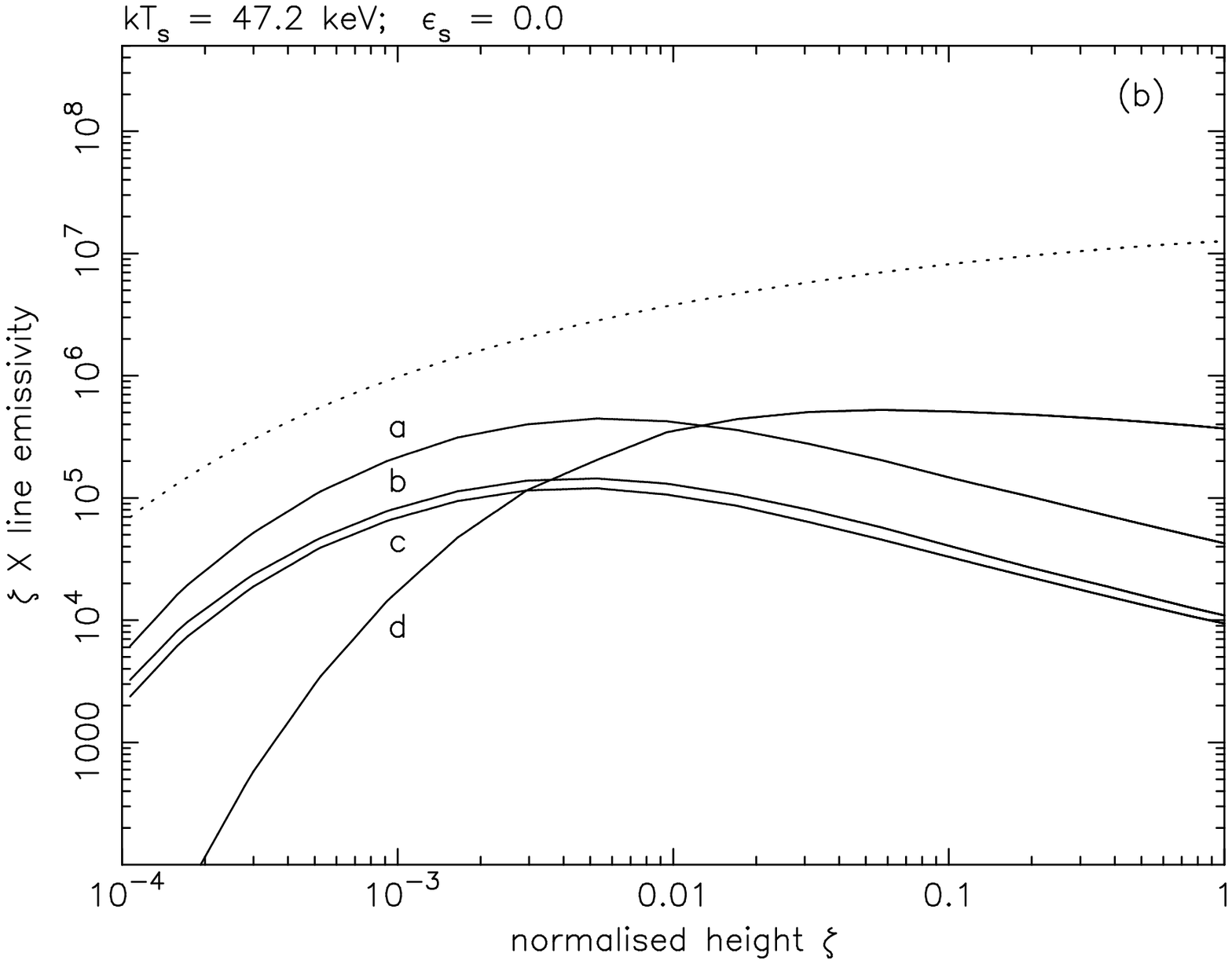}          
\caption{
  (a) The ``$\zeta\times P_{l}$'' emissivity profile of the continuum at 6.8~keV (dotted curve), 
     the Fe~XXV He4 line (curve a), the Fe~XXV He5 line (curve b), 
     the Fe~XXV He6 line (curve c) 
     and the Fe~XXVI Lyman-$\alpha$ line (curve d) 
  for a 0.5-M$_\odot$ strongly magnetised white dwarf 
     with $\epsilon_{\rm s} =$ 100. 
  (b) Same as (a) for a 1.0-M$_\odot$ white dwarf 
     with $\epsilon_{\rm s} =$ 0. }  
\label{fig.10}     
\end{figure}  

In Figures 3 and 4 we show the relative ion concentrations of Iron 
  as a function of the height above the white-dwarf surface 
  for the 1.0-M$_\odot$ and  0.5-M$_\odot$ white dwarfs respectively. 
The normalisation is such that 
  the total concentration is 1.0 at the shock surface ($\zeta = 1$).  
As the density 
  (and the electron number density $n_{\rm e}$) increases  
  when $\zeta$ decreases, 
  the peak concentration of each ion species is always $\geq 1.0$. 
Since the temperature and density in the post-shock region
  are monotonic functions of $\zeta$ 
  (if the electrons and ions have the same temperature 
  and the variation of gravity with height is negligible), 
  when the temperature is specified, the density is uniquely determined. 
In ionisation equilibrium, 
  the peak concentration of the ion species depends only on $T_{\rm e}$. 
Thus, for a fixed white-dwarf mass, 
  the peak concentrations occur at the same $T_{\rm e}$ and $n_{\rm e}$, 
  in spite of the different values of $\epsilon_{\rm s}$.   

As shown in Fig 3, Fe~XXVII dominates 
  in most of the shock-heated region of the 1.0-M$_\odot$ white dwarf 
  with $\epsilon_{\rm s} = 0$. 
The lower ionisation states are important 
  only near the low-temperature bottom of the post-shock region. 
For larger $\epsilon_{\rm s}$, the emission region is cooler, 
  and so the lower ionised species becomes more abundant. 
When  $\epsilon_{\rm s} = 100$, the relative concentration of Fe~XXVI 
  is higher than that of Fe~XXVII in the region 
  where $\zeta \sim 0.1 - 0.3$. 
However, as a whole, Fe~XXVII still dominates in the post-shock region.   
For a 0.5-M$_\odot$ white dwarf, 
  the shock temperature $kT_{\rm s} \approx$ 13.6~keV, 
  which is insufficient to completely ionise all the Iron to Fe~XXVII. 
For $\epsilon_{\rm s} \LS 1$ the dominant species is Fe~XXVI, 
  and for $\epsilon_{\rm s} \GS 10$, Fe~XXV. 
As in the previous  case, the lower ionisation states
  becomes more important at the bottom of the post-shock region. 

In Figure 5, we show the mean charge $(\bar z)$ profiles. 
For the 1.0-M$_\odot$ white dwarfs, 
  ${\bar z} \approx$ 25.8 at the shock surface,  
  which implies that Fe~XXVII is the most abundant ion. 
As $\zeta$  decreases, $\bar z$ first decreases slowly to $\approx 24$ 
  and then falls more rapidly afterwards. 
For $\epsilon_{\rm s} = 0$ (bremsstrahlung cooling only), 
  $\bar z \approx 23.2$ at $\zeta = 10^{-4}$; 
  while for $\epsilon_{\rm s} = 100$ 
  (cyclotron cooling dominated), 
  $\bar z \approx 18$ at $\zeta = 10^{-4}$. 
For the 0.5-M$_\odot$ white dwarfs, $\bar z \approx 25.3$ 
  (i.e.\ Fe~XXVI dominates) at $\zeta = 1.0$. 
Similarly to the 1.0-M$_\odot$ case, 
  $\bar z$ decreases slowly with $\zeta$
  until  it reaches $\approx 24$ and then rapidly afterwards. 
For $\epsilon_{\rm s} =$ 0, $\bar z = 17$ at $\zeta \sim 10^{-4}$; 
  and for $\epsilon_{\rm s} =$ 100, $\bar z = 17$ at $\zeta \sim 10^{-3}$. 

\subsection{Line emissivity profile} 

In Figures 6, 7, 8 and 9, the line emissivity  
  of the Fe~XXVI Lyman-$\alpha$ (1s $^2$S -- 2p $^2$P), 
  Fe~XXV He4 (1s$^2$ $^1$S -- 1s2p $^1$P), 
  Fe XXV~He5 (1s$^2$ $^1$S -- 1s2p $^3$P), 
  and Fe~XXV He6 (1s$^2$ $^1$S -- 1s2p $^3$S) lines are shown. 
The corresponding line centre energies of these lines 
  are 6.93, 6.70, 6.68  and 6.64~keV. 
In calculating the emissivity we have interpolated 
  the table of line power 
  given in  Mewe, Gronenschild \& van den Oord (1985).  
The electron number density that we assume 
  corresponds to an specific accretion rate  
  of $\dot m =$ 1~g~cm$^{-2}$~s$^{-1}$. 

When the white-dwarf mass and the accretion rate are fixed, 
  the emissivity of an emission line is the same at the shock 
  for all $\epsilon_{\rm s}$. 
As $\epsilon_{\rm s}$ increases, cyclotron cooling becomes more efficient. 
As the electron temperature of the shock-heated region is lowered, 
  the lower ionised species becomes more significant. 
As a result, the emissivities of the Fe~XXVI and Fe~XXV lines 
  all peak at larger $\zeta$ for larger $\epsilon_{\rm s}$ 
  (Fig.~10).   

\section{Discussion} 

\subsection{White-dwarf mass determination} 
  
When the local emissivity 
  $P_{\rm l}(\zeta;T_{\rm s},\epsilon_{\rm s},z_{\rm i})$ 
  of an emission line $z_{\rm i}$ is specified, 
the intensity of the line is simply the sum of contribution 
  from all zones in the emission region, i.e.,      
\begin{equation} 
 I(z_{\rm i})\ \approx \ {{x_{\rm s} S}\over {4\pi D^2}} 
    A(Z)\ \int^1_{\Delta_{\rm i}} d\zeta 
    P_{\rm l}(\zeta;T_{\rm s},\epsilon_{\rm s},z_{\rm i})\ ,   
\end{equation} 
  where $A(Z)$ is the abundance of the element that 
  give rise to the emission line $z_{\rm i}$, 
  $\Delta_{\rm i}$ the height above the white-dwarf surface 
  at which the emission becomes optically thick, 
  $S$ the cross-section area of the emission region, 
  and $D$ is the distance of the  source. 
In terms of the temperature gradient $dT/d\zeta$, 
\begin{equation} 
 I(z_{\rm i})\ =\ {{x_{\rm s} S}\over {4\pi D^2}} 
    A(Z)\ T_{\rm s} \int^{T_{\rm s}}_{{T_{\Delta_{\rm i}}}} dT 
    {{d \zeta}\over{d T}}
    P_{\rm l}(\zeta;T_{\rm s},\epsilon_{\rm s},z_{\rm i})\ ,    
\end{equation} 
   where $T_{\Delta_{\rm i}}$ is the temperature at $\Delta_{\rm i}$. 
The ratio of the intensities of two emission lines 
   $z_{\rm i}$ and $z_{\rm j}$ 
   of the same element $Z$ are therefore 
\begin{equation} 
 R(z_{\rm i},z_{\rm j};T_{\rm s},
     \epsilon_{\rm s},\Delta_{\rm i},\Delta_{\rm j})\  =\ 
  {{\int^{T_{\rm s}}_{{T_{\Delta_{\rm i}}}} dT {{d \zeta}\over {d T}}
  P_{\rm l}(\zeta;T_{\rm s},\epsilon_{\rm s},z_{\rm i})} \over 
  {\int^{T_{\rm s}}_{{T_{\Delta_{\rm j}}}} dT {{d \zeta}\over {d T}}
  P_{\rm l}(\zeta;T_{\rm s},\epsilon_{\rm s},z_{\rm j})}} \ .       
\end{equation} 
As the line-intensity ratio is a function of   
  $\epsilon_{\rm s}$, $T_{\rm s}$, 
  $T_{\Delta_{\rm i}}$ and $T_{\Delta_{\rm j}}$ only,    
  one may use the observed  line intensity ratio 
  to constrain these parameters.

\begin{figure} 
\centering 
\epsfxsize=8.2cm       
\epsfbox{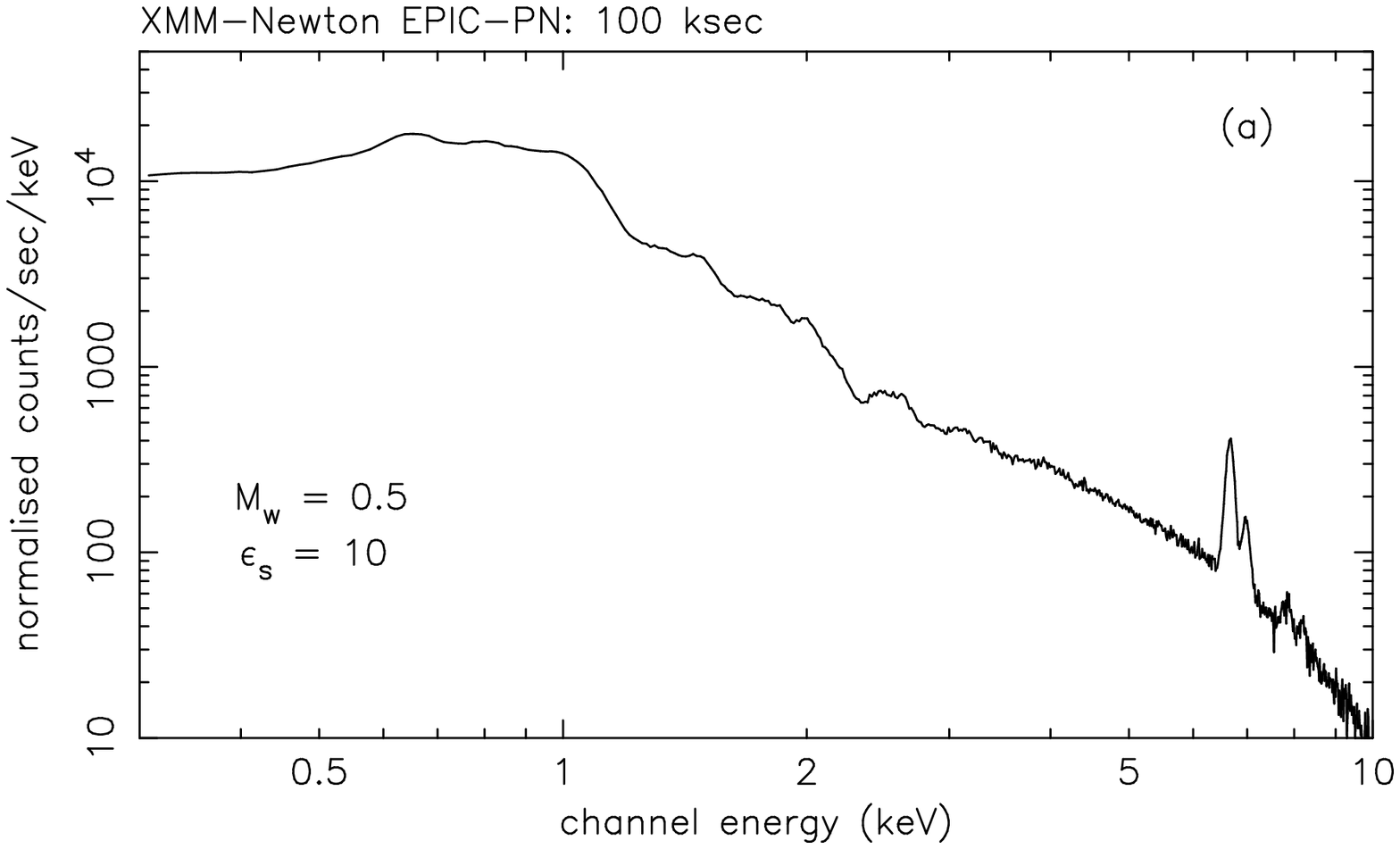}  
\epsfxsize=8.2cm    
\epsfbox{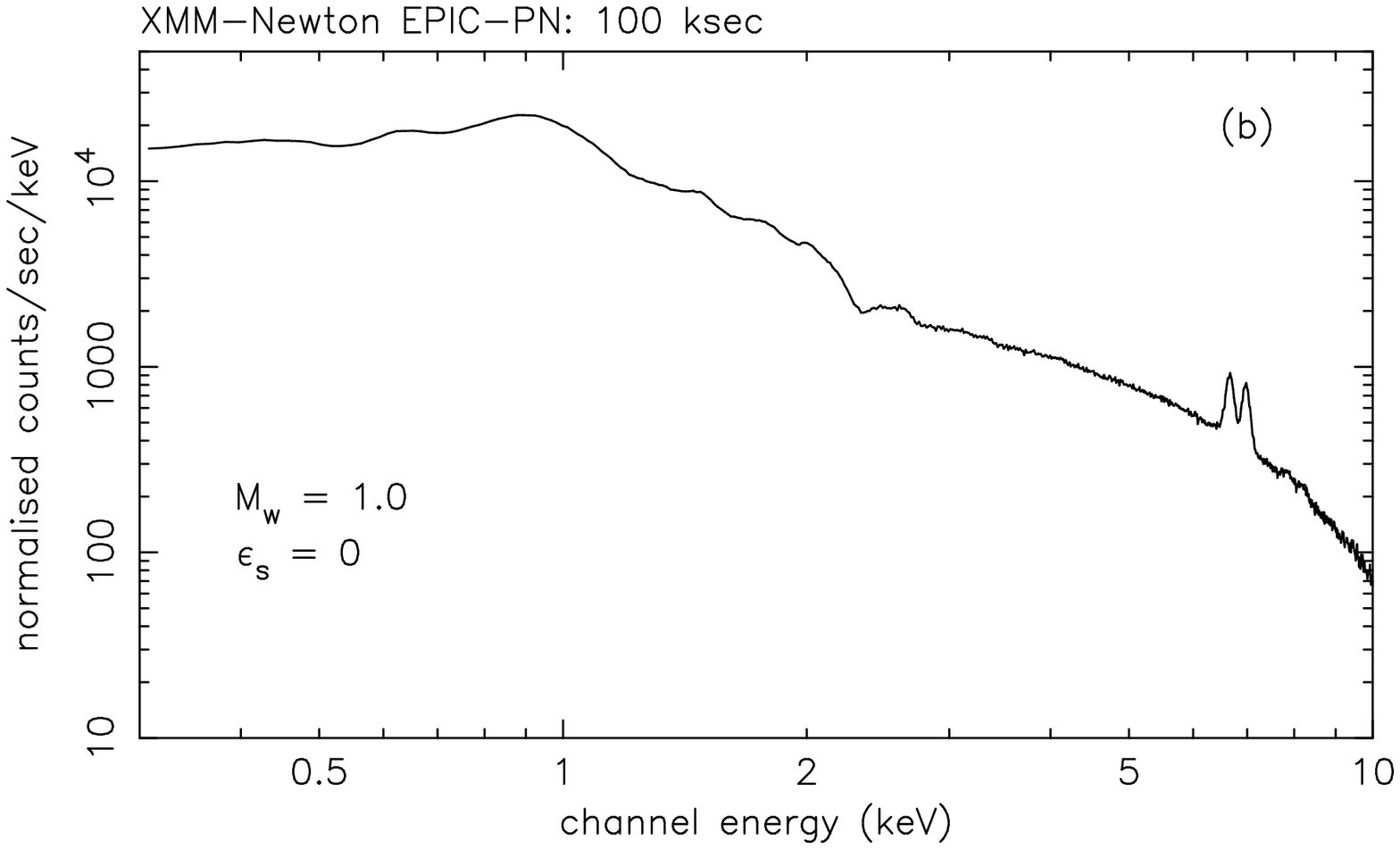}          
\caption{(a) Simulated {\it XMM-Newton} EPIC-PN spectra 
   of a 0.5-M$_\odot$ white dwarf 
   with substantial cyclotron cooling ($\epsilon_{\rm s} = 10$) 
   for a 100-ks exposure. We assume solar metal abundance. 
   The flux is normalised 
   to that of the observed flux of AM Her itself in the 2--10keV
   energy band
   ($\approx 7 \times 10^{-11}~{\rm erg}~{\rm s}^{-1}~{\rm cm}^{-2} $) 
   derived from {\it ASCA} data (0.5--10keV) (Ishida et al.\ 1997).  
   (b) Same as (a) for a 1.0-M$_\odot$ white dwarf with no cyclotron cooling. 
   In simulating the spectra we have used calibration files 
   derived 7 months into the orbit of {\it XMM-Newton} 
   and assume a medium filter.}  
\label{fig.11 }     
\end{figure}      

Fujimoto (1996) and Fujimoto \& Ishida (1995, 1997) assumed that 
  $T_{\Delta_{\rm j}} = T_{\Delta_{\rm i}} (= T_\Delta)$, 
  and use the ratio of H- and He-like Fe K$\alpha$ lines 
  observed by {\it ASCA} to determine the shock $T_{\rm s}$ 
  of the Intermediate Polar EX~Hya. 
From the deduced shock temperature they obtained 
  a white-dwarf mass of 0.48$^{+0.10}_{-0.06}$~M$_\odot$. 
Using the same method, Hellier et al.\ (1996) deduced 
  the white-dwarf mass of another Intermediate Polar AO~Psc, 
  giving 0.40$^{+0.07}_{-0.11}$~M$_\odot$.   
These masses are generally in agreement with those  
  determined by fitting the X-ray continuum using {\it Ginga} data  
  (0.45~M$_\odot$ and 0.45~M$_\odot$ respectively
  for EX~Hya and AO~Psc)  
  and {\it RXTE} data  
  (0.44$\pm0.03$~M$_\odot$ and 0.60$\pm0.03$~M$_\odot$ respectively) 
  (Cropper, Wu \& Ramsay 1999; Ramsay 2000). 

As the emissivity of the emission lines is sensitive 
  to the temperature of the emission region, 
  it is generally considered that white-dwarf masses 
  can be accurately determined using spectral lines,  
  and the line emission may provide better constraints 
  to the white-dwarf masses than the continuum. 
However, an accurate measurement of the line strengths 
  does not necessarily lead to an accurate determination 
  of the {\it shock} temperature.  
The shock temperature can be deduced 
  only if there are sufficient concentrations of ions 
  that are responsible for the line emission at the shock. 
If the shock temperature is sufficiently high 
  that the element is completely ionised at the shock,  
  the strengths of its H- and He-like lines 
  are insensitive to the shock temperature.  
These lines are now emitted from the cooler bottom 
  of the post-shock region (cf.\ Fig.~10a and Fig.~10b).  
Thus, using Fe H- and He-like emission lines 
  to determine white-dwarf mass is practical 
  for the low-mass ($\sim 0.5$~M$_\odot$) systems 
  but not for massive white dwarfs (Wu 2000).  
Indeed, an attempt to constrain the white-dwarf masses 
  of two more massive AM~Her type systems, 
  AM~Her itself and BL~Hyi, 
  using this method was unsuccessful (Fujimoto \& Ishida 1995). 
Although one may consider elements with higher ionisation potential, 
  such as Nickel, 
  the abundance of heavier elements in typical mCVs 
  are generally low and unable to produce lines 
  that are strong enough to be useful in typical observations. 
For instance in Figures~11, 12, 13 and 14 we show 
  the strengths of the Fe and other lines 
  in simulated {\it XMM-Newton} EPIC-PN and RGS spectra  
  of accreting white dwarfs of different masses and cooling efficiencies. 
In these simulations we have used calibration files determined 
  7 months into flight 
  and used exposures typical of {\it XMM-Newton} observations. 
Tennant et al.\ (1998) shows simulated spectra using the 
  {\it Chandra} grating spectra.

\begin{figure} 
\centering 
\epsfxsize=8.0cm       
\epsfbox{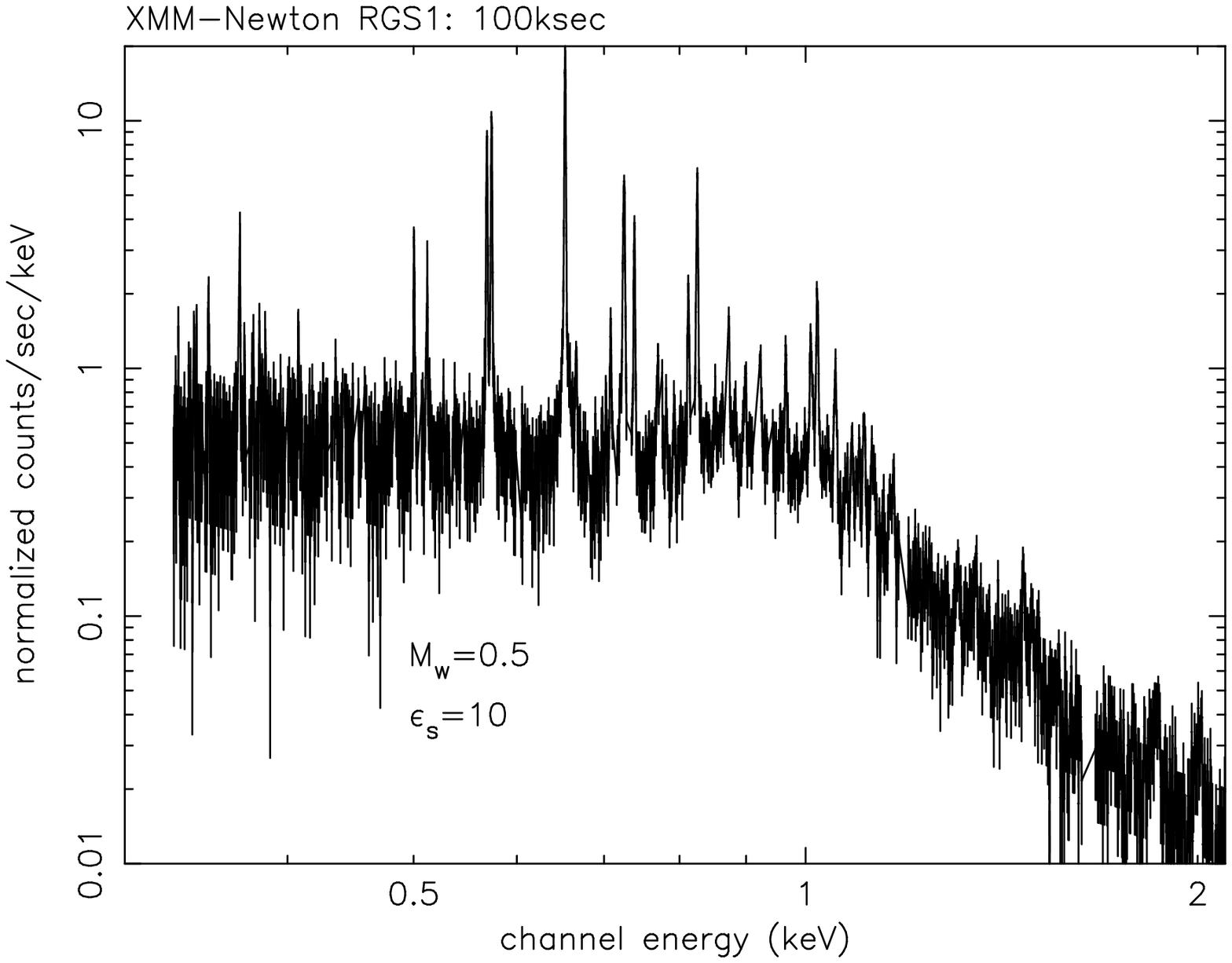}  
\epsfxsize=8.0cm  
\epsfbox{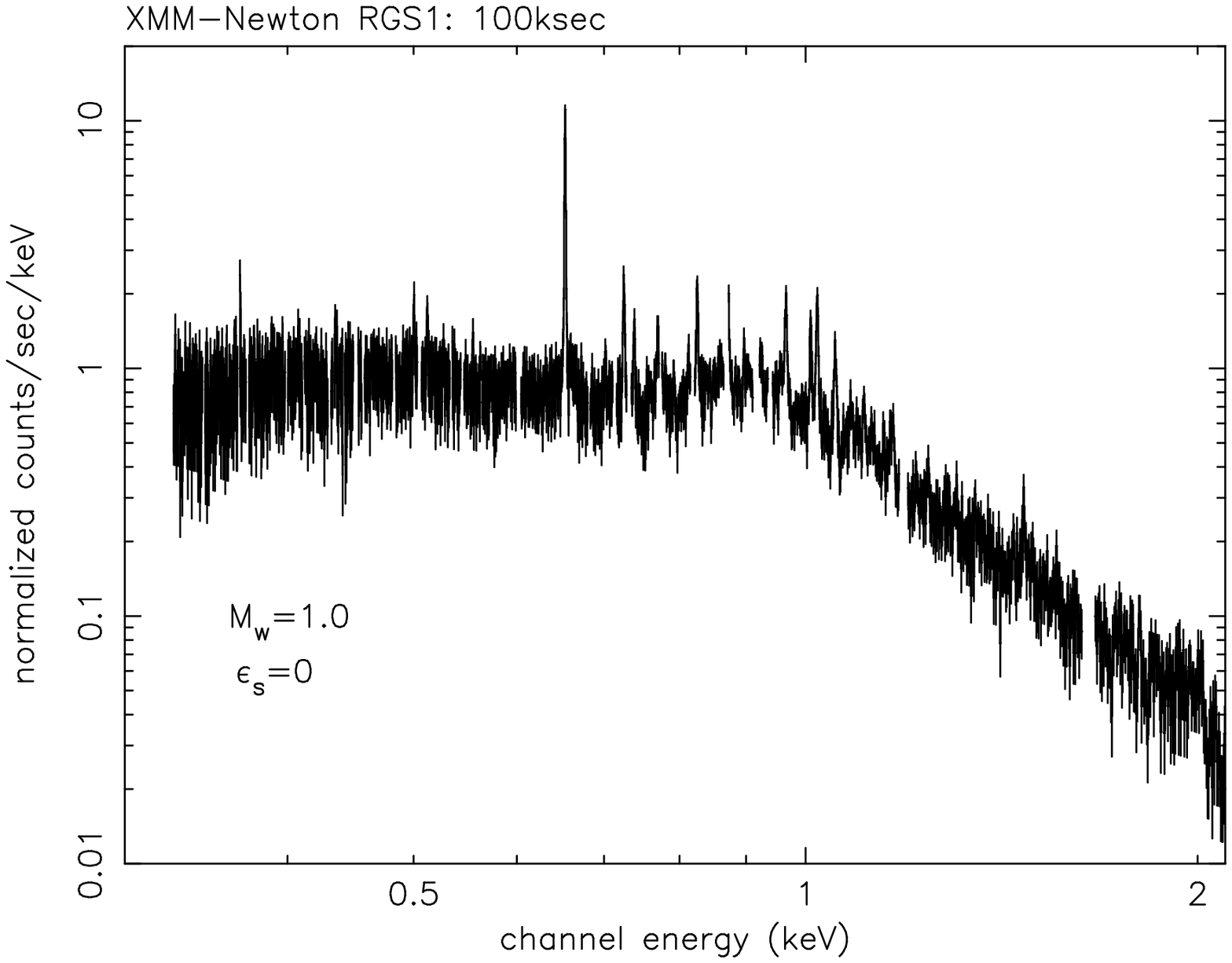}          
\caption{(a) Simulated {\it XMM-Newton} RGS spectra (0.3--2.1keV)
      of a 0.5-M$_\odot$ white dwarf 
      with substantial cyclotron cooling ($\epsilon_{\rm s} = a0$) 
      for a 100-ks exposure.  
   We assume solar metal abundance. 
   The flux is normalised to the observed flux 
      of the hard component of AM Her itself     
      ($\approx 1.7\times 10^{-11}~{\rm erg}~{\rm s}^{-1}~{\rm cm}^{-2} $)     
      determined from {\it ROSAT} data (0.1--2.4keV) 
      (Ramsay et al.\ 1994).      
  (b) Same as (a) for a 1.0-M$_\odot$ white dwarf with no cyclotron cooling. 
  In simulating the spectra we have used calibration files 
     derived 7 months into the orbit of {\it XMM-Newton}. }   
\label{fig.12}     
\end{figure}    

\subsection{Flow diagnosis}  

Because of the temperature and velocity stratification 
  in the shock-heated region, 
  the emission lines from different heights 
  above the white-dwarf surface have different Doppler shifts. 
A line can be emitted only if the temperature 
  of the accretion matter 
  allows the according electronic transition to occur, 
  so one can infer the temperature and density of the region 
  where the line is emitted. 
In addition the Doppler shift of the line centre energy 
  can be measured, 
  from which the line-of-sight velocity of the emitter 
  can be deduced. 
Thus, by examining the different lines in a spectrum, 
  one can obtain a relation 
  between the flow velocity, temperature and density.    

For a 1.0-M$_\odot$ white dwarf, the velocity of material  
  at the accretion shock is about $2\times 10^8$~cm~s$^{-1}$. 
The centre energy of lines emitted from regions 
  near the shock will have an energy shift of about 0.6\%. 
As the white dwarf in a mCV revolves with the orbit 
  and it also rotates, 
  there is an additional velocity shift 
  of about $2-3 \times 10^7$~cm~s$^{-1}$ for typical CV  parameters. 
The white-dwarf rotational period and the binary orbital period 
  are the same for typical AM Her type systems, 
  and it allows us to subtract the velocity 
  due to orbital motion and the white-dwarf rotation easily. 
Moreover, AM Her type systems do not have an accretion disc, 
  and therefore the line spectra are emitted mainly 
  from the hot shock-heated accretion matter 
  near the white-dwarf surface. 
The Doppler shift of the emission lines 
  is then only due to the bulk accretion flow, 
  and they can be extracted as described above 
  from the orbital phase-binned data. 

A line-centre energy resolution of 1 part in 10000 
  (achievable by {\it XMM-Newton} and {\it Chandra} at about 1~keV) 
  will allow us to measure the accretion flow 
  a few tens metres above the white-dwarf surface. 
As shown in Figures~13 and 14, 
  the Fe complex and other lines are clearly resolved 
  in the {\it XMM-Newton} simulated grating spectra 
  for a 100-ksec observation. 
In the parallel study of Tennant et al.\ (1998),  
  it was shown that an energy resolution of 0.1-eV can be achieved 
  by  the Medium Energy Gratings of {\it Chandra}  
  for a single gaussian fit to a line of centre energy of 1~keV 
  for about 200 photons in the lines.  
For a mCV of the X-ray brightness of the system AM Her itself, 
  this requires an observation of about 100~ksec 
  to detect this number photons in 1/10 of an orbital phase. 

X-ray spectroscopy will therefore allow the flow velocity  
  in these accreting systems to be measured directly. 
This diagnostic power cannot be achieved by optical spectroscopy,   
  despite its better velocity resolution, 
   as optical emission from  the shock-heated material 
   in mCVs is optically thick. 

\section{Summary}   
 
We have investigated 
  the ionisation structure and line emissivity profiles 
  of the shock-heated emission regions of accreting white dwarfs 
  in mCVs by means of an analytic model 
  given in Wu (1994) and Wu, Chanmugam \& Shaviv (1994). 
We have found that 
  for Iron, the corona-condition approximation is generally satisfied 
  in most of the shock-heated region where lines are emitted. 
Because of the temperature and density stratification 
  in the shock-heated region, 
  different lines are emitted from different heights 
  above the white-dwarf surface. 
By measuring the Doppler shifts of the lines, 
  one can obtain a relationship between the flow velocity, 
  the temperature and the density of the emitting gas, 
  thus providing a means to map the hydrodynamics 
  of the post-shock accretion flow directly. 
Our study also indicates that using emission lines 
  to determine white-dwarf masses is practical only  
  for low-mass systems. 
For massive systems, the abundant elements 
  such as Fe are completely ionised at the shock 
  and so their line emission is insensitive to the shock temperature. 

\section*{Acknowledgement} 
 
We thank Allyn Tennant for discussions 
  and the referee, Manabu Ishida,  
  for the suggestions to improve the manuscript. 
KW acknowledges the support of an ARC Australian Research Fellowship
  and a PPARC visiting fellowship.

\end{document}